\documentclass[journal]{IEEEtran}

\usepackage{cite}

\ifCLASSINFOpdf
  \usepackage[pdftex]{graphicx}
  \graphicspath{{./figures/}}

\else

\fi

\usepackage{subcaption}

\usepackage{float}
\usepackage{amsmath}
\usepackage{amssymb}
\usepackage{amstext}
\usepackage{amsfonts}
\usepackage{url}
\usepackage{bbm}
\usepackage{color}

\usepackage{tablefootnote}

\usepackage{float}

\usepackage{color}
\definecolor{red}{RGB}{255,0,0}

\usepackage{multirow}

\usepackage{epstopdf}
\usepackage{graphicx}
\usepackage[section]{placeins}
\usepackage{xspace}
\usepackage{booktabs}
\usepackage{verbatim}

\newcommand{\eg}{\textit{e.g.}\xspace}
\newcommand{\etal}{\textit{et al.}\xspace}

\usepackage[linesnumbered,lined,noend,ruled]{algorithm2e}

\begin{document}
%
\title{Local Intensity Order Transformation for Robust Curvilinear Object Segmentation}

\author{Tianyi Shi,
        \and Nicolas Boutry,
        \and Yongchao Xu,
        \and Thierry Géraud
        \thanks{ T. Shi is with the School of EIC, Huazhong University of Science and Technology (HUST), Wuhan, 430074, China (E-mail: shitianyihust@hust.edu.cn).}
        \thanks{N. Boutry and T. G\'eraud are with Laboratoire de Recherche et D\'eveloppement de l'EPITA (LRDE), Le Kremlin-Bicêtre, 94276, France (E-mail: \{nicolas.boutry, thierry.geraud\}@lrde.epita.fr).}
        \thanks{Y. Xu is with the School of Computer Science, Wuhan University, Wuhan, 430070, China (E-mail: yongchao.xu@whu.edu.cn).}
\thanks{\textit{Corresponding author: Yongchao Xu.}}
}

\markboth{Journal of \LaTeX\ Class Files,~Vol.~XX, No.~XX, August~2021}%
{Shell \MakeLowercase{\textit{et al.}}: Bare Demo of IEEEtran.cls for IEEE Journals}
\maketitle

\begin{abstract}

Segmentation of curvilinear structures is important in many applications, such as retinal blood vessel segmentation for early detection of vessel diseases and 
pavement crack segmentation for road condition evaluation and maintenance.
Currently, deep learning-based methods have achieved impressive performance on these tasks. Yet, most of them mainly focus on finding powerful deep architectures but ignore capturing the inherent curvilinear structure feature (\eg, the curvilinear structure is darker than the context) for a more robust representation. In consequence, the performance usually drops a lot on cross-datasets, which poses great challenges in practice. In this paper, we aim to improve the generalizability by introducing a novel local intensity order transformation (LIOT). Specifically, we transfer a gray-scale image into a contrast-invariant four-channel image based on the intensity order between each pixel and its nearby pixels along with the four (horizontal and vertical) directions.
This results in a representation that preserves the inherent characteristic of the curvilinear structure while being robust to contrast changes. Cross-dataset evaluation on three retinal blood vessel segmentation datasets demonstrates that LIOT improves the generalizability of some state-of-the-art methods.
Additionally, the cross-dataset evaluation between retinal blood vessel segmentation and pavement crack segmentation shows that LIOT is able to preserve the inherent characteristic of curvilinear structure with large appearance gaps.
An implementation of the proposed method is available at \url{https://github.com/TY-Shi/LIOT}.

\end{abstract}

\IEEEpeerreviewmaketitle

\section{Introduction}
\label{sec:introduction}
Curvilinear structures often appear in biomedical analysis. Their segmentation is very important in applications like retinal fundus disease screening~\cite{abramoff2010retinal,fraz2012blood}, early detection of vessel diseases, and in biometric authentication systems~\cite{liu2011fingerprint, laibacher2019m2u, su2019non}. In parallel, curvilinear structures also appear in pavement crack segmentation, which is useful for road condition evaluation and road maintenance~\cite{zou2012cracktree}. 


Compared to general object segmentation, curvilinear structure segmentation faces some particular challenges~\cite{bibiloni2016survey}: 1) thin, long, and tortuosity shapes; 2) inadequate contrast between curvilinear structures and the surrounding background; 3) uneven background illumination; 4) various image appearances.
To cope with these challenges, classical curvilinear object segmentation methods~\cite{koller1995multiscale,carlotto2006enhancement,soares2006Gabor_waveletretinal,ricci2007linedetector,al2009active,marin2010new,obara2012contrast,xiao2012multiscale,zou2012cracktree,krylov2014stochastic,annunziata2015boosting,vicas2015detecting,turetken2016reconstructing,merveille2017curvilinear,strisciuglio2019robust,merveille2019n} mainly focus on designing specifically engineered features. They usually adopt filters or some morphological tools to capture one or more specific features from images. Such hand-crafted feature-based methods usually require careful parameter tuning, which is difficult to handle a wide variety of complex curvilinear structure segmentation.






\begin{figure}
\centering
\begin{subfigure}[b]{0.32\linewidth}
\centering
\includegraphics[width=1.0\linewidth]{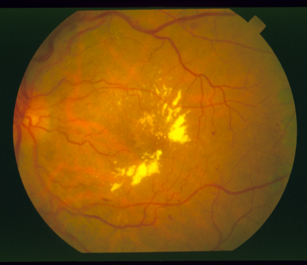}
\end{subfigure}
\begin{subfigure}[b]{0.32\linewidth}
\centering
\includegraphics[width=1.0\linewidth]{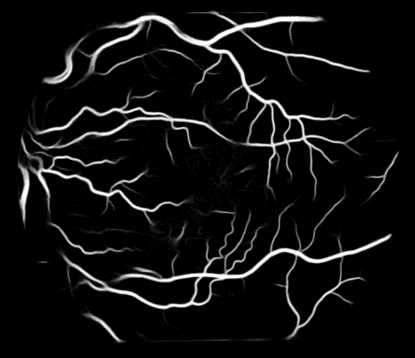}
\end{subfigure}
\begin{subfigure}[b]{0.32\linewidth}
\centering
\includegraphics[width=1.0\linewidth]{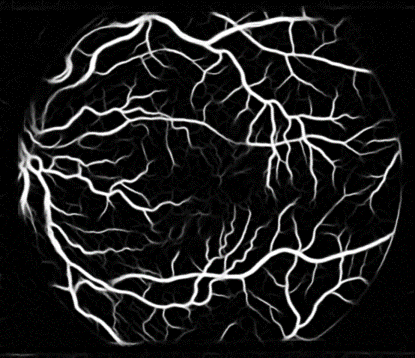}
\end{subfigure}
\vskip 0.2cm
\begin{subfigure}[b]{0.32\linewidth}
\centering
\includegraphics[width=1.0\linewidth]{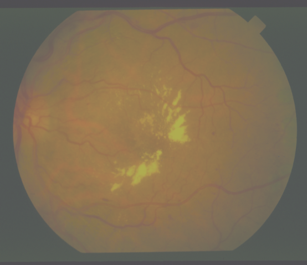}
\caption{Input}
\end{subfigure}
\begin{subfigure}[b]{0.32\linewidth}
\centering
\includegraphics[width=1.0\linewidth]{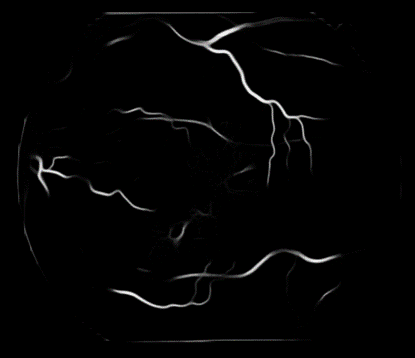}
\caption{Seg. on input}
\end{subfigure}
\begin{subfigure}[b]{0.32\linewidth}
\centering
\includegraphics[width=1.0\linewidth]{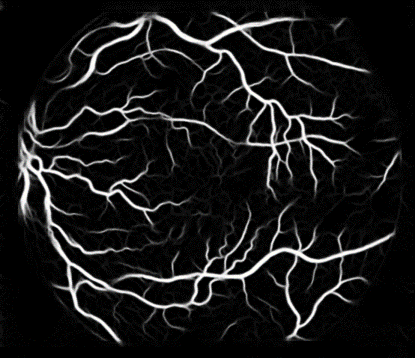}
\caption{Seg. on LIOT}
\end{subfigure}
\caption{An example of segmentation results on an image and the same image with changed contrast. Top row: segmentation on the original image by IterNet~\cite{li2019iternet}; Bottom row: segmentation on contrast changed image. Different from direct segmentation on the original image, the proposed LIOT is invariant to contrast changes and thus yields robust segmentation results.}
\label{fig:example}
\end{figure}

Most recent methods~\cite{li2015cross,maninis2016DRIU,mosinska2018beyond,yan2018joint,luan2018gabor,mosinska2019joint,wang2019context,mou2019cs,guo2019bts,cherukuri2019deep,dey2020subpixel,mou2020cs2,ding2020novel,zhang20203d,nazir2020off,wu2021scs,shen2021modeling} leverage deep learning~\cite{lecun2015deep, he2020hybrid, yue2021self, xie2021super} for curvilinear object segmentation and have achieved significant improvement over previous methods. These methods mainly focus on designing various network architectures or loss functions to improve the segmentation performance. Despite the high in-dataset accuracy for the state-of-the-art methods, they usually do not generalize well to images with different image appearances. For instance, as illustrated in Fig.~\ref{fig:example}, one may achieve good results on the image similar to the training dataset. The segmentation accuracy greatly drops on the same image with slightly changed contrast. Yet, the inherent characteristic of curvilinear structure (\eg, darker than the context) holds for both images. This implies that the generalizability of curvilinear object segmentation methods still remains challenging. Designing a robust method that captures the inherent property of curvilinear structure is of great interest.

In this paper, we aim to address the above issue and focus on improving the generalizability of current deep learning-based methods in segmenting curvilinear objects. Instead of directly operating on the original image, we introduce a novel image transformation called \emph{Local Intensity Order Transformation (LIOT)}, which is a representation dedicated to curvilinear structure and invariant to the increasing change of image contrast. Specifically, we compare the intensity order of a given pixel to the values of its nearby neighbors (\textit{e.g.} along with the four horizontal and vertical directions within a certain distance). In this way, we transform the original image into a new four-channel one, which does not depend on the absolute intensity value but the relative intensity order to capture the inherent characteristic of the curvilinear structure. 
An example of segmentation on LIOT is illustrated in Fig.~\ref{fig:example}. LIOT is robust to contrast changes, yielding very similar segmentation result for the original image and the one with contrast change.

The main property of the proposed LIOT is the robustness to curvilinear characteristic and contrast change. LIOT simply changes the input and can be plugged into any deep learning-based method. Since the main goal of this paper is not to develop a powerful pipeline that outperforms all other methods in segmenting curvilinear objects, we simply apply LIOT to the recent method IterNet~\cite{li2019iternet} and conduct cross-dataset experiments on four datasets: DRIVE~\cite{DRIVE2004ridge}, STARE~\cite{STARE2000locating}, CHASEDB1~\cite{fraz2012ensemble}, and CrackTree~\cite{zou2012cracktree}. These cross-dataset experiments demonstrate that the proposed contrast-insensitive LIOT is simple yet effective to help deep learning-based methods to capture the characteristic of the curvilinear structure, and improves the generalizability of existing methods.

The main contribution of this paper is as follows: 1) We propose a novel image transform method that captures the inherent characteristic of curvilinear structures and is invariant to the increasing change of image contrast; 2) The spirit of using LIOT as input to the deep learning-based method enables the model to express a better generalization and to be more robust to non-targeted universal adversarial perturbations.

The rest of this paper is organized as follows. We shortly review some related works in Section~\ref{sec:relatedworks}. The proposed method is then detailed in Section~\ref{sec:method}, followed by extensive experimental results in Section~\ref{sec:experiments}. Finally, we conclude and give some perspectives in Section~\ref{sec:conclusion}.

\section{Related work}
\label{sec:relatedworks}

Curvilinear structure segmentation has been widely exploited recently. We first review some representative classical and deep learning-based methods for curvilinear object segmentation in Section~\ref{subsec:segmentation methods}. A review of some related works on leveraging intensity order-based information, in particular variants of census transform and intensity order-based feature description, are discussed in Section~\ref{subsec:local transform}. The comparison of the proposed LIOT with some related works is depicted in Section~\ref{subsec:comparison}. 

\subsection{Curvilinear object segmentation}
\label{subsec:segmentation methods}

Before the era of deep learning, curvilinear object segmentation pipelines~\cite{koller1995multiscale,carlotto2006enhancement,soares2006Gabor_waveletretinal,ricci2007linedetector,al2009active,marin2010new,obara2012contrast,xiao2012multiscale,zou2012cracktree,krylov2014stochastic,annunziata2015boosting,vicas2015detecting,turetken2016reconstructing,merveille2017curvilinear,strisciuglio2019robust,merveille2019n} are usually based on different techniques ranging from hand-crafted filters to machine learning approaches. 
Most recent methods shift to deep neural networks to directly learn effective features from the training data. Some of these methods are categorized and detailed in the following.

\medskip
\noindent
\textbf{Classical methods:} 
Classical curvilinear object segmentation methods mainly rely on extracting engineered features. For instance, Koller \textit{et al.}~\cite{koller1995multiscale} introduce a method based on a non-linear combination of linear filters, with an edge-detection approach. Carlotto \textit{et al.}~\cite{carlotto2006enhancement} combine directional filter to enhance low-contrast curvilinear. 
In~\cite{soares2006Gabor_waveletretinal}, the authors leverage pixel intensity and 2D Gabor wavelet transform for retinal blood vessel segmentation. In~\cite{ricci2007linedetector}, the authors develop the line operator to construct a feature vector for supervised classification using the support vector machine. 
AI-Diri \textit{et al.}~\cite{al2009active} segment vessels in retinal images with a related active contour model.
Mar\`in \textit{et al.}~\cite{marin2010new} utilize a 7-D vector for pixel representation. Xiao \textit{et al.}~\cite{xiao2012multiscale} propose to replace the low-level Gaussian kernel with a bi-Gaussian to detect curvilinear structures. The orientation analysis of gradient vector fields has been proposed for retinal blood vessel segmentation in~\cite{fraz2012ensemble}. 
Obara \textit{et al.}~\cite{obara2012contrast} propose a contrast-independent approach to identify curvilinear structures based on oriented phase congruency.
Krylov \textit{et al.}~\cite{krylov2014stochastic} develop a stochastic approach for line segment extraction.
Vicas \textit{et al.}~\cite{vicas2015detecting} use a structure tensor to extract meaningful low-level information for curvilinear structures.
In~\cite{turetken2016reconstructing}, the authors rely on graph-based representations for reconstructing curvilinear networks.
Merveille \etal~\cite{merveille2017curvilinear} and ~\cite{wang2020higher} segment curvilinear structures by ranking the orientation responses of operators. Strisciuglio \textit{et al.}~\cite{strisciuglio2019robust} propose a novel operator for the delineation of curvilinear structures in images. Merveille \textit{et al.}~\cite{merveille2019n} propose a mixed gradient operator for segmentation task.

\medskip 
\noindent
\textbf{Deep learning-based methods:} Deep learning-based methods learn features from the training data and significantly improve the performance of curvilinear object segmentation. For example, the neural network is used for retinal blood vessel segmentation in~\cite{li2015cross}, which does not depend on the carefully engineered feature. Maninis \textit{et al.}~\cite{maninis2016DRIU} propose a unified framework of retinal image analysis that jointly performs both retinal blood vessel and optic disc segmentation. In~\cite{yan2018joint}, Yan \textit{et al.} propose a joint segment-level and pixel-wise loss for retinal blood vessel segmentation. Luan \textit{et al.}~\cite{luan2018gabor} develop the Gabor convolution networks to reinforce the robustness of learned features against orientation and scale changes. Guo \textit{et al.}~\cite{guo2019bts} present a multi-scale deeply supervised network with short connections, achieving good cross-dataset results. 
In~\cite{cherukuri2019deep}, Cherukuri \textit{et al.} propose a deep model with regularization under geometric priors. Mosinska \textit{et al.}~\cite{mosinska2019joint} propose a multi-task learning of curvilinear structure segmentation and path classification. Wang \textit{et al.}~\cite{wang2019context} propose a context-aware spatio-recurrent network for segmenting curvilinear structure. 
Lei \textit{et al.}~\cite{mou2019cs,mou2020cs2} propose a channel and spatial attention network based on U-Net, effectively extracting curvilinear structures from three biomedical imaging modalities. Ma \textit{et al.}~\cite{ma2021rose} introduce a novel coarse-to-fine vessel segmentation network with the ability to separately detect thick and thin vessels. Dey~\cite{dey2020subpixel} introduces a subpixel convolution and a feature fusion technique for retinal blood vessel segmentation. The distance field of tubular objects~\cite{wang2019tubular,wang2020deep} has been proposed to improve the segmentation performance. Some recent methods~\cite{mosinska2018beyond,hu2019topology,hu2021topology,shit2021cldice,wang2021single} aim to improve the topological mistakes by proposing topology-aware loss for curvilinear object segmentation. The method in~\cite{cheng2021joint} focuses on developing powerful network architecture for curvilinear object segmentation by leveraging boundary as a geometric constraint to refine feature. 
\begin{figure}
\centering
\begin{subfigure}[b]{0.24\linewidth}
\centering
\includegraphics[width=1.0\linewidth]{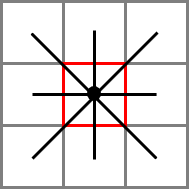}
\caption{CT~\cite{zabih1996non}}
\end{subfigure}
\begin{subfigure}[b]{0.24\linewidth}
\centering
\includegraphics[width=1.0\linewidth]{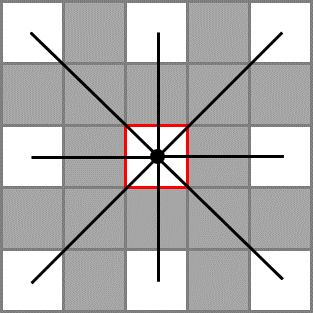}
\caption{SP-CT~\cite{chang2010algorithm}}
\end{subfigure}
\begin{subfigure}[b]{0.24\linewidth}
\centering
\includegraphics[width=1.0\linewidth]{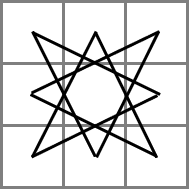}
\caption{SCT~\cite{lee2016improved}}
\end{subfigure}
\begin{subfigure}[b]{0.24\linewidth}
\centering
\includegraphics[width=1.0\linewidth]{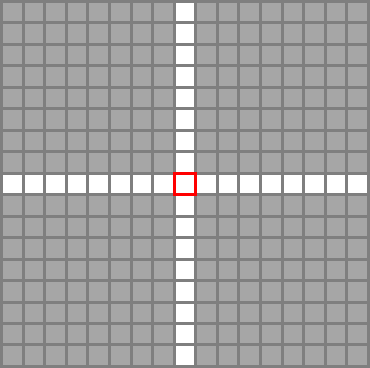}
\caption{LIOT}
\end{subfigure}
\caption{Comparison of LIOT with variants of census transform, which compare the intensity order between white pixels and the red one in the center for CT and Sparse CT (SP-CT).}
\label{fig:censuscomparison}
\end{figure}

\subsection{Census transform and intensity order-based feature description}
\label{subsec:local transform}
Census Transform (CT)~\cite{zabih1996non} is a non-parametric local transform, which is originally defined as an ordered set of comparisons of pixel intensity in local $3 \times 3$ neighborhood.
Froba \textit{et al.}~\cite{froba2004face} propose the MeanCT that leverages the mean intensity of the neighborhood to increase the robustness.
Ambrosch \textit{et al.}~\cite{ambrosch2010miniature} propose a SparseCT which uses a subset of pixels within the region of a larger neighborhood.
Chang \textit{et al.}~\cite{chang2010algorithm} also propose a sparse CT Mini-census. In~\cite{fife2012improved}, Fife \textit{et al.} propose generalized census transform (GCT) to define a family of masks in a $5 \times 5$ neighborhood with different levels of sparsity. Similarly, the center symmetric CT (CSCT)~\cite{spangenberg2013weighted} compares a pair of pixels within the census window. The star census transform (SCT)~\cite{lee2016improved} extends GCT by defining masks of symmetrical sequences of connected edges, of equal length, forming star-shaped scan-patterns around the center.
Ahlberg \textit{et al.}~\cite{ahlberg2019unbounded} propose a genetic algorithm to find a new and powerful Census Transform method. 
Yu \textit{et al.}~\cite{yu2020stereo} adopt the census transform to calculate the illumination characteristics for tree reconstruction optimization. Lai \textit{et al.}~\cite{lai2019efficient} also use census transform in medical imaging.
Ram{\'\i}rez  \textit{et al.}~\cite{cornejo2019audio} apply the census transform operator for audio-visual emotion recognition, owing to its robustness to monotonic changes. 

Apart from the variants of census transform, some works~\cite{wang2011local,wang2015exploring,fan2011rotationally,shen2014hep,lei2014local,wang2017ordinal} also share the similar idea with the proposed LIOT on leveraging the local intensity order information, but with different purposes (local feature description~\cite{wang2011local,wang2015exploring,fan2011rotationally} and image classification~\cite{shen2014hep,lei2014local,wang2017ordinal}).

\subsection{Comparison with related works}
\label{subsec:comparison}

\textbf{LIOT Versus Classical methods:} Classical methods rely on engineered features to extract curvilinear structure. Therefore, these methods usually require careful parameter tuning and are hard to generate good performance in a wide variety of complex curvilinear object segmentation. The proposed LIOT can leverage a deep convolutional neural network (CNN) to learn more effective features.
LIOT can be plugged into any CNN architecture for curvilinear object segmentation.

\textbf{LIOT Versus Deep learning-based methods:} Most deep learning-based methods for curvilinear object segmentation are mainly inspired by recent CNN-based object/image segmentation methods. They usually focus on improving the in-dataset segmentation accuracy by improving feature learning, and ignore the limited cross-dataset performance due to the appearance gap between different datasets. This hinders the transferability to other similar curvilinear object segmentation. The proposed LIOT aims to capture the inherent characteristic of curvilinear objects, and thus to improve the generalization performance in various curvilinear object segmentation for deep learning-based methods.


\textbf{LIOT Versus Census transform and intensity order-based feature description:} The proposed LIOT is in spirit similar to the census transform, which is a non-parametric transform that depends on the relative intensity ordering instead of absolute intensity values. This makes them invariant to increasing contrast changes. As shown in Fig.~\ref{fig:censuscomparison}, compared with variants of census transform, the proposed LIOT is specifically designed for the curvilinear object that captures the curvilinear structure from four directions.
Compared with the simple intensity order comparison between the center pixel and its surrounding neighborhood pixels for census transform, the horizontal and vertical long-range pixels capture better the inherent property of curvilinear objects having usually thin and long structures.
The LIOT also differs with the intensity order-based feature description methods in how to define the pairs of pixels for intensity order comparison.
Besides, to the best of our knowledge, we are the first to combine intensity order information and deep-learning for robust curvilinear structure segmentation, which enables the model to express a better generalization and to be more robust to non-targeted universal adversarial perturbations.

\begin{figure}[t]
\includegraphics[width=\linewidth]{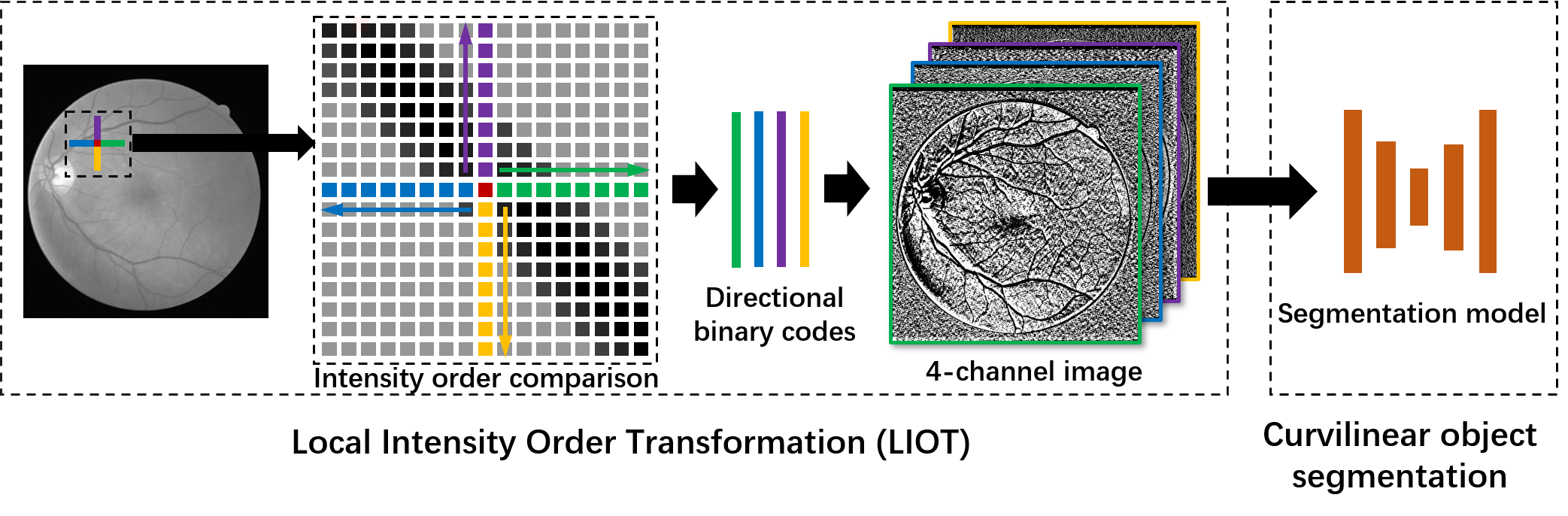}
\caption{Overview of the proposed LIOT for curvilinear object segmentation. We transform an input image to a 4-channel intensity-order-based image, which is then fed into a segmentation network.}
\label{overview}
\end{figure}


\section{Proposed Method}\label{sec:method}

\begin{figure*}[t]
\centering
\includegraphics[width=0.8\linewidth]{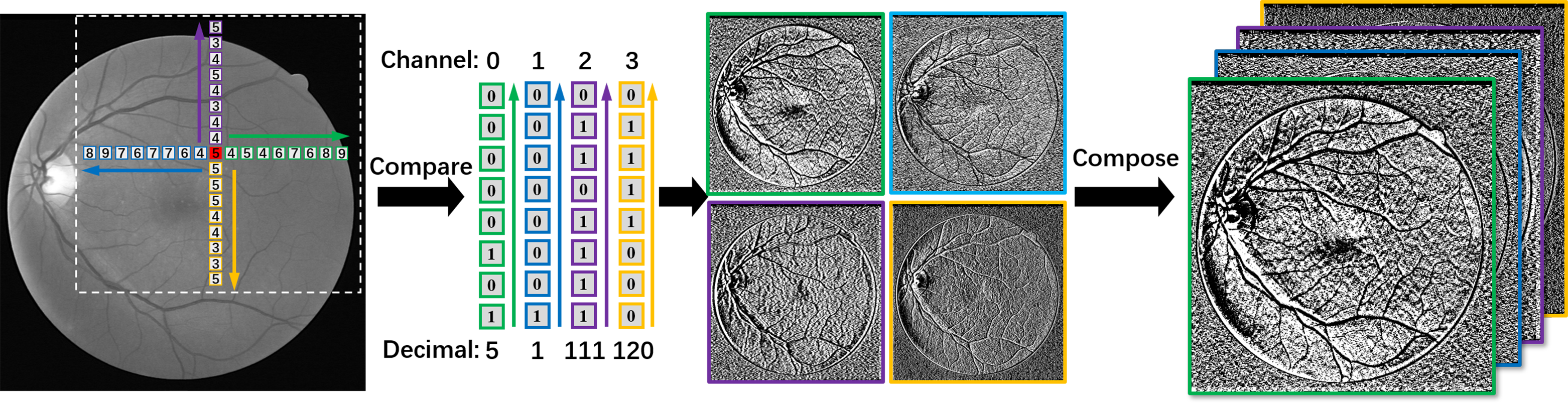}
\caption{Illustration of LIOT. For each pixel $p$, we compare its intensity with the value of every group of 8 neighboring pixels lying perpendicular with $p$, resulting in a 4-channel image with values in each channel ranging from 0 to 255 (8 bits).}
\label{algorithm}
\end{figure*}

\subsection{Overview}
\label{subsec:overview}
Curvilinear objects have entirely different color distributions and backgrounds for different application scenarios. Such an appearance gap makes it difficult for the algorithm to accurately segment different curvilinear objects in different images. Thus, a general method to improve the generalizability for curvilinear segmentation algorithms is required. For that, 
we introduce a novel image transformation LIOT based on the relative intensity order between each pixel and its perpendicular neighbors. Such a transformation does not depend on the absolute value of each pixel, and is thus more robust to contrast changes. 
The overall pipeline of the proposed method is depicted in Fig.~\ref{overview}. We first convert the given image into a gray-scale one. Then we rely on the local intensity order to compute four directional binary codes, forming a 4-channel image that captures the curvilinear structure characteristic. We then feed this contrast invariant 4-channel image into a segmentation network.


\subsection{Local intensity order transformation}
\label{subsec:liot}

Though different curvilinear object images may have various contrast and intensity distribution, the curvilinear structure is always darker (without loss of generality, we can invert the image when the curvilinear structure is brighter) than the context. Based on this property, we propose a novel and robust representation called LIOT that converts an input gray-scale image into a 4-channel intensity-order-based image. More precisely, as illustrated in Fig.~\ref{algorithm}, for each pixel $p$ in the domain $\Omega$ of an image $f$, we compare the value of $f(p)$ with each group of 8 neighboring pixels $\{n_s^i \, | \, i = 1, \dots, 8\}$ (with Euclidean distance to $p$ ranging from 1 to 8) lying perpendicular to $p$, where $s \in \{l, r, t, b\}$ denotes left, right, top, and bottom side of $p$, respectively. This results in four 2D 8-bit images, where the $i$-th bit of each converted 2D image $f'_s$ on $p$ corresponds to the binary code given by the intensity order between $f(p)$ and $f(n_s^i)$. These four 2D images of directional binary codes are concatenated together, composing a 4-channel intensity-order-based image $F = [f'_l, f'_r, f'_t, f'_b]$. Formally, for each direction side $s \in \{l, r, t, b\}$, we compute the corresponding image $f'_s$ of directional binary codes on each pixel $p$ as follows:
\begin{equation}
    f'_s(p) \, = \, \sum_{i = 1}^8 [f(p) > f(n_s^i)] \times 2^{i-1},
    \label{eq:liot}
\end{equation}
where $[f(p) > f(n_s^i)]$ is 1 if the value $f(p)$ is larger than  $f(n_s^i)$, otherwise 0.

As depicted in Eq.~\eqref{eq:liot}, the proposed transformation is irrespective of absolute intensity values, and only depends on the relative intensity order. LIOT captures the ``darker than'' property of curvilinear structure and is invariant to increasing change of contrast. This makes LIOT more robust to appearance differences for images from different applications. Therefore, LIOT can improve the generalizability of curvilinear object segmentation methods.
\subsection{Network architecture}
\label{subsec:networkarchitecture}

Since the major goal is to improve the generalizability of existing CNN-based methods, we do not aim to develop a powerful network that outperforms other methods. Instead, we simply apply LIOT with the recent IterNet~\cite{li2019iternet} by changing its input channel to the 4-channel image given by LIOT. Iternet is a U-Net like encoder-decoder model that adopts U-Net as the base module and combines three mini U-Nets. Each mini U-Net uses feature maps from its precedent module with fewer parameters than U-Net.
IterNet also adds skip-connections from the base U-Net to all mini U-Nets and the connections among the mini U-Nets. We simply change the first convolution layer channel from 3 to 4-channel for adapting LIOT inputs, and keep the following layers the same as the original IterNet~\cite{li2019iternet}.

\subsection{Training objective}
\label{subsec:trainingobjective}

We leverage the network depicted in Section~\ref{subsec:networkarchitecture} to segment the curvilinear object. 
The loss is the same as that in IterNet~\cite{li2019iternet}.
The network parameters are optimized with the sigmoid cross-entropy loss, defined as:
\begin{equation}
L \; = \; \sum_i\big(-y_{i}log(p_{i})-(1-y_{i})log(1-p_{i})\big),
\label{eq:loss}
\end{equation}
where $y_{i}$ represents the binary indicator (0 or 1) whether the pixel $i$ belongs to the ground-truth curvilinear object, and $p_{i}$ is the predicted probability that the pixel $i$ is a foreground pixel.

During the testing phase, we convert the original gray-scale image to a 4-channel LIOT image, and feed LIOT into the segmentation network to compute the final probability map.



\section{Experiments}\label{sec:experiments}

We evaluate the proposed method on four widely adopted datasets: DRIVE~\cite{DRIVE2004ridge}, STARE~\cite{STARE2000locating},  CHASEDB1~\cite{fraz2012ensemble}, and CrackTree~\cite{zou2012cracktree}. 
A short description of these datasets and adopted evaluation protocol are given in Section~\ref{subsec:datasets}. Some implementation details are then depicted in Section~\ref{subsec:implementationdetails}. 
To demonstrate the generalizability of the proposed LIOT, we conduct four types of evaluations: 1) In-dataset evaluation in Section~\ref{subsec:in-dataset compare} to demonstrate that the proposed LIOT does not significantly decrease the performance on original images; 2) In-dataset evaluation on images with universal adversarial perturbations (see Section~\ref{subsec:in-dataset-perturbation}) to show that LIOT is more robust to universal adversarial attacks; 3) Cross retinal dataset evaluation in Section~\ref{subsec:Cross-retinal}, showing that LIOT generalizes better to images with small domain gap; 4) Evaluation on cross-dataset between Retinal and CrackTree dataset (see Section~\ref{subsec:Cross-crack}), proving that LIOT is more robust to large domain changes. In Section~\ref{subsec:moreillustrations}, we also show some results of applying LIOT trained on the retinal dataset to images with very different types of curvilinear structures, further demonstrating the generalizability of LIOT.


\subsection{Datasets and evaluation protocol}
\label{subsec:datasets}
\noindent\textbf{DRIVE}~\cite{DRIVE2004ridge}: The DRIVE dataset contains the curvilinear-shaped vessel. This dataset consists of 40 $565 \times 584$ color retinal images, which are split into 20 training images and 20 test images.

\medskip

\noindent\textbf{STARE}~\cite{STARE2000locating}: The STARE dataset consists of 20 $700 \times 605$ color retina images divided into 10 training and 10 test images.

\medskip

\noindent\textbf{CHASEDB1}~\cite{fraz2012ensemble}: The CHASEDB1 dataset is composed of 28 $999 \times 960$ color retinal images, which are split into 20 training images and 8 test images.

\medskip

\noindent\textbf{CrackTree}~\cite{zou2012cracktree}:  The CrackTree dataset contains 206 $800 \times 600$ pavement images with different kinds of cracks having curvilinear structure. The whole dataset is split into 160 training and 46 test images. Following~\cite{mosinska2018beyond}, we dilate the annotated centerlines by 4 pixels to form the ground-truth segmentation.  
As shown in Fig.~\ref{crossCrack_retinalresult}, the multiple shadows and cluttered background make the segmentation a challenging task. 


\medskip


Images with curvilinear objects may have different image resolutions, and thus have varied thickness (in terms of pixels) of curvilinear structure. To cope with such a scale gap, for the retinal images, we first resize each image to a similar scale based on the size of the field of view (FOV) and image size. Specifically, we keep the image size of DRIVE unchanged, and resize the images in STARE, CHASEDB1 accordingly. Precisely, we resize STARE images from $700 \times 605$ to $554 \times 479$, and CHASEDB1 images from $999 \times 960$ to $584 \times 561$. The images in the CrackTree dataset are resized to $512 \times 512$. These size settings are used for both baseline methods and the proposed method. For retinal images, since LIOT requires a total order between pixel values, we thus convert the resized color image to a gray-scale one by selecting its green channel.



\medskip
\noindent\textbf{Evaluation protocol}: Following the classical evaluation protocol for curvilinear object segmentation, we first compute the true positives (TPs), false positives (FPs), false negatives (FNs), and true negatives (TNs) for the segmentation result.
Then, the classical accuracy (Acc), sensitivity (Se), specificity (Sp), area under the receiver operating characteristics curve (AUC), and F1-score are used to evaluate the performance. Besides, we also adopt the connectivity~\cite{evaluation2011function}: $1 - \min\big(1, |\#_C(S_G) - \#_C(S)|/\#(S_G)\big)$, where $\#_C(S_G)$ and $\#_C(S)$ denotes the number of connected components in the ground-truth and segmented result, respectively. $\#(S_G)$ stands for the number of pixels in the ground-truth segmentation. This connectivity metric reflects the curvilinear structural continuity and is useful for branching analysis, further assessing the effectiveness of the proposed LIOT. We compute the metrics in the field of view area for retinal datasets: DRIVE, STARE , and CHASEDB1. For the CrackTree dataset, the metrics are computed on the whole image.
For all methods on all involved datasets, we report the results by using the optimal segmentation threshold based on the F1-score.


\subsection{Implementation details}
\label{subsec:implementationdetails}
We adopt classical data augmentation strategy to increase the training data and avoid over-fitting. Specifically, we randomly rotate images from -180 to 180 degrees, and shear from -0.1 to 0.1. We also flip images with the horizontal and vertical direction. Then, images are randomly shifted from -0.1 to 0.1, and randomly zoomed from 0.8 to 1.2. Finally, we randomly crop the augmented images to $128 \times 128$.

Since the major goal is not to develop a powerful network that outperforms other methods, we simply apply LIOT with the recent IterNet~\cite{li2019iternet}. Following~\cite{li2019iternet}, we employ a batch size of 32 to train the network using cross-entropy loss for 1000 epochs. To further demonstrate the usefulness of the proposed LIOT, we also conduct experiments by using the topological loss (Topo)~\cite{hu2019topology}. We adopt Adam~\cite{kingma2014adam} with a learning rate of 0.001 to optimize the network.
For Topo~\cite{hu2019topology} method, we set different optimal weights for the topological loss term on different datasets: 0.005 for DRIVE, 0.001 for CHASEDB1 and STARE, and 0.01 for CrackTree. For census~\cite{zabih1996non} method, we choose the window size 3$\times$3 as the original census definition and do not set any other parameters because the census is a non-parametric local transformation. 
During inference, overlapping image patches are extracted with a stride equal to 8. 

\setlength{\tabcolsep}{2.3pt}
\begin{table}[t]
\caption{Quantitative in-dataset evaluation of LIOT compared with the baseline model and Census transform.}\label{indatasettab}
\centering
\begin{tabular}{|c|c|c|c|c|c|c|c|}
\hline
In-Dataset                              & Methods             & Se    & Sp      & Acc      & AUC      & F1     &Connectivity\\ \hline
\multirow{5}{1.7cm}{\centering CrackTree\\$\Rightarrow$\\ CrackTree} 
                                          & Baseline~\cite{li2019iternet}  & \bf{0.813}& 0.995 & 0.992   & \bf{0.994}   & 0.754 &0.851\\
                                          & Census~\cite{zabih1996non}  & 0.493& 0.996 & 0.989   & 0.898   & 0.566 &0.563\\  
                                          & LIOT & 0.780 & \textbf{0.997}  & \textbf{0.994}   & 0.990 & \bf{0.790} &\bf{0.916}\\ 
                                          \cline{2-8}
                                          & Topo~\cite{hu2019topology}  & \textbf{0.848}& 0.995 & 0.993   & \textbf{0.996}   & 0.769 &0.870\\
                                          & Topo+LIOT  &0.802 & \textbf{0.997}   & \textbf{0.994}   & 0.988 &\textbf{0.801} &\textbf{0.917}\\ 
                                          \hline
\multirow{5}{1.7cm}{\centering DRIVE\\$\Rightarrow$\\ DRIVE}
                                          & Baseline~\cite{li2019iternet}  & \textbf{0.828}& 0.973 & \textbf{0.954}   & \bf{0.978}   & \bf{0.821} & \bf{0.875}\\
                                          & Census~\cite{zabih1996non}  & 0.783 & 0.972 & 0.948   & 0.962   & 0.794 &0.831\\
                                          & LIOT & 0.811 &\textbf{0.974}  &0.953   &0.976 &0.814 &0.763\\ 
                                          \cline{2-8}
                                          & Topo~\cite{hu2019topology}  & \textbf{0.828}& \bf{0.973} & \textbf{0.954}   & \textbf{0.980}   & \textbf{0.822} &\textbf{0.878}\\
                                          & Topo+LIOT  & 0.819& 0.972 & 0.953   & 0.977   & 0.815 &0.777\\
                                          \hline
\multirow{5}{1.7cm}{\centering STARE \\$\Rightarrow$\\ STARE}  
                                          & Baseline~\cite{li2019iternet}  &0.812& \textbf{0.982}  & \bf{0.964}   & \bf{0.984}   & \bf{0.826} & \bf{0.886}\\
                                          & Census~\cite{zabih1996non}  & 0.763& 0.975 & 0.953   & 0.964   & 0.771 &0.641\\ 
                                          & LIOT & \bf{0.816} & 0.977   & 0.960   & 0.983    & 0.810 &0.803\\
                                          \cline{2-8}
                                          & Topo~\cite{hu2019topology}  & \textbf{0.836}& \bf{0.979} & \textbf{0.964}   & \textbf{0.985}   & \textbf{0.829} &\textbf{0.887}\\
                                          & Topo+LIOT  & 0.817& 0.976 & 0.959   & 0.982   & 0.807 &0.803\\
                                          \hline
\multirow{5}{1.7cm}{\centering CHASEDB1 \\$\Rightarrow$\\ CHASEDB1}  
                                          & Baseline~\cite{li2019iternet}  &\bf{0.833}& \textbf{0.978}  & \textbf{0.965}   & \textbf{0.986}   & \bf{0.811} &0.715\\
                                          & Census~\cite{zabih1996non}  & 0.738& 0.979 & 0.957   & 0.963   & 0.756 &0.682\\ 
                                           & LIOT & 0.816 & 0.976   & 0.961   & 0.983    & 0.792 &\bf{0.803}\\
                                           \cline{2-8}
                                          & Topo~\cite{hu2019topology}  & \textbf{0.838}& \textbf{0.978} & \textbf{0.965}   & \textbf{0.986}   & \textbf{0.812} &0.735\\
                                          & Topo+LIOT  & 0.819& 0.977 & 0.963   & 0.983   & 0.799 &\textbf{0.829}\\ 
                                          \hline\hline                                         
\multirow{5}{1.7cm}{\centering Average}  
                                          & Baseline~\cite{li2019iternet}  &\bf{0.822}& \textbf{0.982}  & \textbf{0.969}   & \bf{0.986}   & \bf{0.803} & \bf{0.832}\\
                                          & Census~\cite{zabih1996non}  & 0.694& 0.980& 0.962   & 0.947   & 0.722 &0.679\\ 
                                          & LIOT &0.807 & 0.981   & 0.967   & 0.983    & 0.801 & 0.821\\
                                          \cline{2-8}                                            
                                          & Topo~\cite{hu2019topology}  & \textbf{0.838}& \bf{0.981} & \textbf{0.969}   & \textbf{0.987}   & \textbf{0.808} &\textbf{0.843}\\
                                          & Topo+LIOT  & 0.814& \bf{0.981} & 0.967   & 0.983   & 0.806 &0.832\\                                           
                                          \hline
\end{tabular}
\end{table}

\setlength{\tabcolsep}{2.3pt}
\begin{table}[t]
\caption{Quantitative in-dataset evaluation on images with non-targeted universal adversarial perturbation~\cite{poursaeed2018generative} of LIOT compared with the baseline model.}\label{indatasetPerturbationtab}
\centering
\begin{tabular}{|c|c|c|c|c|c|c|c|}
\hline
In-Dataset                              & Methods             & Se    & Sp      & Acc      & AUC      & F1     &Connectivity\\ \hline
\multirow{5}{1.7cm}{\centering CrackTree\\$\Rightarrow$\\ CrackTree} 
                                          & Baseline~\cite{li2019iternet}  & 0.453& 0.993 & 0.985   & 0.963   & 0.474 &0.342\\
                                          & Census~\cite{zabih1996non}  &0.564&\bf{0.996} &0.990   &0.952   &0.613 &0.677\\
                                          & LIOT & \textbf{0.668} & \textbf{0.996}  & \textbf{0.992}   & \textbf{0.976} & \bf{0.699} &\bf{0.788}\\ 
                                          \cline{2-8}
                                          & Topo~\cite{hu2019topology}  &0.439&\bf{0.995} &0.987   &0.950   &0.496 &0.270\\
                                          & Topo+LIOT  &\bf{0.625} &\bf{0.995}   &\bf{0.990}   &\bf{0.971}&\bf{0.641} &\bf{0.748}\\ 
                                          \hline
\multirow{5}{1.7cm}{\centering DRIVE\\$\Rightarrow$\\ DRIVE}
                                          & Baseline~\cite{li2019iternet}  & 0.729 &0.842 & 0.828   & 0.871   & 0.518 & 0.611\\
                                          & Census~\cite{zabih1996non}  &0.591&0.778 &0.755   &0.762   &0.380 &0.621\\
                                          & LIOT & \textbf{0.752} &\textbf{0.967}  &\textbf{0.940}   &\textbf{0.952} &\textbf{0.761} &\textbf{0.714}\\ 
                                          \cline{2-8}
                                          & Topo~\cite{hu2019topology}  & 0.715&0.863 & 0.844   &0.876   &0.538 &0.574\\
                                          & Topo+LIOT  &\textbf{0.738}& \textbf{0.958} & \textbf{0.930}   & \textbf{0.942}   &\textbf{0.728} &\textbf{0.705}\\
                                          \hline
\multirow{5}{1.7cm}{\centering STARE \\$\Rightarrow$\\ STARE}  
                                          & Baseline~\cite{li2019iternet}  &0.429&0.880  &0.833   & 0.754   &0.389 &0.373\\
                                          & Census~\cite{zabih1996non}  &0.420&\bf{0.976} &\bf{0.918}   &\bf{0.842}   &0.518 &0.283\\
                                          & LIOT & \bf{0.461} & 0.965   & 0.912   &0.827    & \bf{0.523} &\bf{0.396}\\
                                          \cline{2-8}
                                          & Topo~\cite{hu2019topology}  & 0.363&0.909 &0.852   &0.740   &0.338 &0.318\\
                                          & Topo+LIOT  &\bf{0.447} &\bf{0.966}   &\bf{0.912}   &\bf{0.813}&\bf{0.514} &\bf{0.345}\\ 
                                          \hline
\multirow{5}{1.7cm}{\centering CHASEDB1 \\$\Rightarrow$\\ CHASEDB1}  
                                          & Baseline~\cite{li2019iternet}  &0.633&0.899 &0.875   &0.867   &0.477 &0.322\\
                                          & Census~\cite{zabih1996non}  &0.781&\bf{0.977} &\bf{0.960}   &0.973   &0.778 &0.763\\
                                          & LIOT  &\bf{0.788} &\bf{0.977}   &\bf{0.960}   &\bf{0.978}&\bf{0.779} &\bf{0.776}\\ 
                                           \cline{2-8}
                                          & Topo~\cite{hu2019topology}  &0.682&0.876 &0.859   &0.873   &0.466 &0.428\\
                                          & Topo+LIOT  &\bf{0.778} &\bf{0.978}   &\bf{0.960}   &\bf{0.978} &\bf{0.779}&\bf{0.776}\\ 
                                          \hline\hline                                         
\multirow{5}{1.7cm}{\centering Average}  
                                          & Baseline~\cite{li2019iternet}  &0.561&0.904 &0.880   &0.864   &0.455 &0.412\\
                                          & Census~\cite{zabih1996non}  &0.589&0.932 &0.906   &0.882   &0.572 &0.586\\                                          
                                          & LIOT  &\bf{0.667} &\bf{0.976}   &\bf{0.951}   &\bf{0.933}&\bf{0.691} &\bf{0.668}\\ 
                                           \cline{2-8}
                                          & Topo~\cite{hu2019topology}  &0.550&0.911 &0.885   &0.860   &0.460 &0.398\\
                                          & Topo+LIOT  &\bf{0.647} &\bf{0.974}   &\bf{0.948}   &\bf{0.926}&\bf{0.665} &\bf{0.644}\\                                            
                                          \hline
\end{tabular}
\end{table}

\subsection{In-Dataset evaluation}
\label{subsec:in-dataset compare}

\begin{figure*}[t]
\centering
\begin{subfigure}[b]{0.80\linewidth}
\centering
\includegraphics[width=1.0\linewidth]{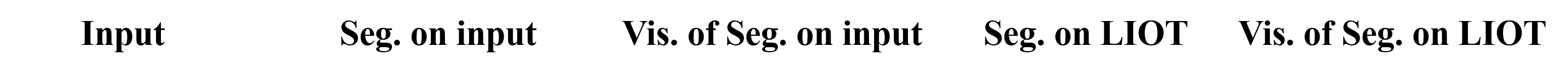}
\end{subfigure}
\centering
\begin{subfigure}[b]{0.80\linewidth}
\centering
\includegraphics[width=1.0\linewidth]{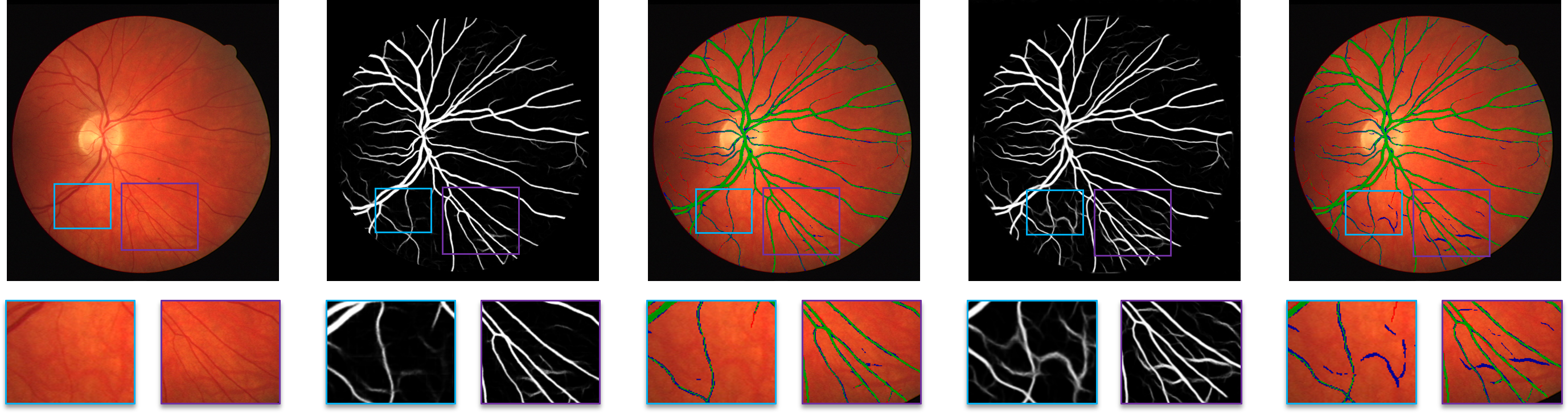}
\caption{In-dataset validation on DRIVE}
\end{subfigure}
\begin{subfigure}[b]{0.80\linewidth}
\centering
\includegraphics[width=1.0\linewidth]{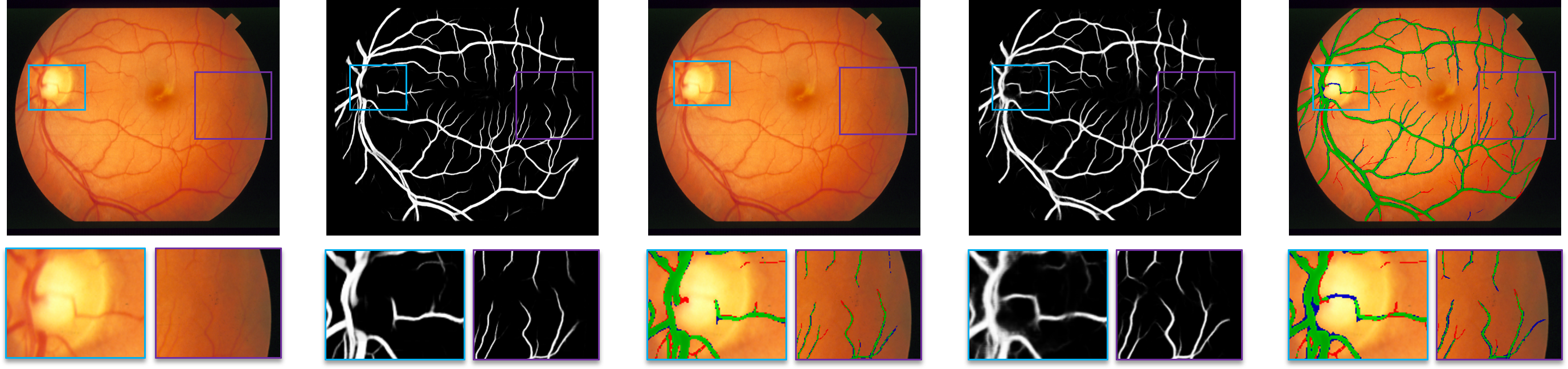}
\caption{In-dataset validation on STARE}
\end{subfigure}
\begin{subfigure}[b]{0.80\linewidth}
\centering
\includegraphics[width=1.0\linewidth]{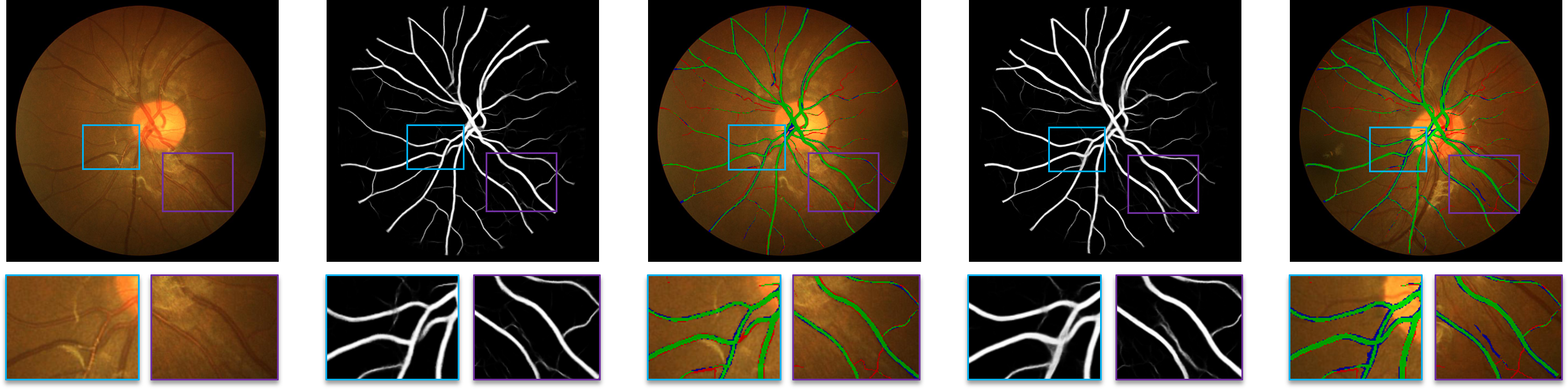}
\caption{In-dataset validation on CHASEDB1}
\end{subfigure}
\begin{subfigure}[b]{0.80\linewidth}
\centering
\includegraphics[width=1.0\linewidth]{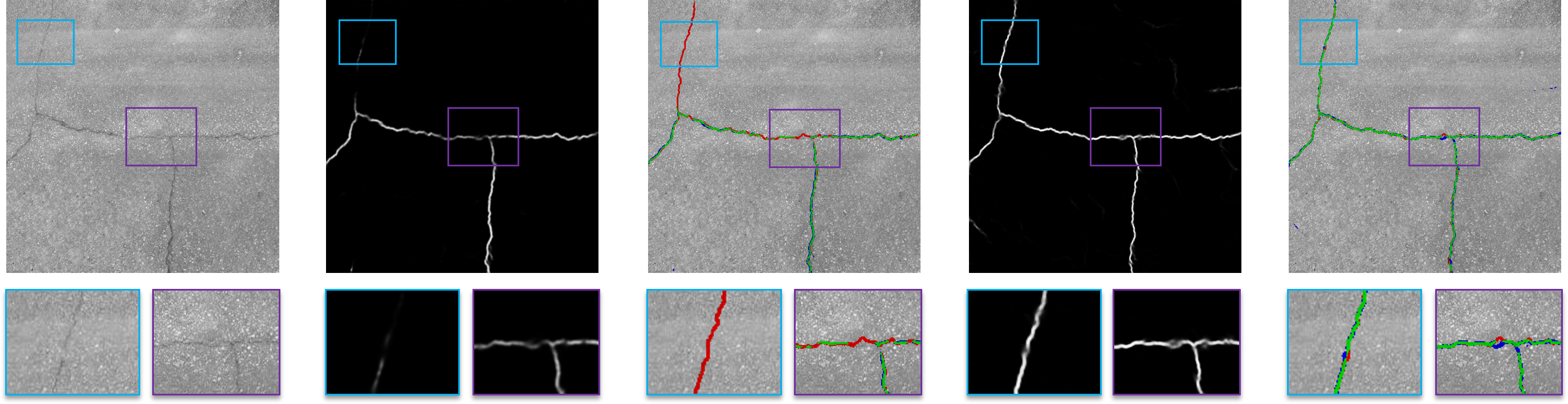}
\caption{In-dataset validation on Cracktree}
\end{subfigure}
\caption{Some segmentation results under in-dataset validation. Green pixels: TPs; Red pixels: FNs; Blue pixels: FPs. Some FPs achieved by LIOT can find evidence in the original image, which might be TPs ignored in the manual annotation. Best viewed by zooming in the electronic version.}
\label{fig:qualitative-indataset}
\end{figure*}

\begin{figure}[t]
\centering
\begin{subfigure}[b]{0.95\linewidth}
\centering
\includegraphics[width=1.0\linewidth]{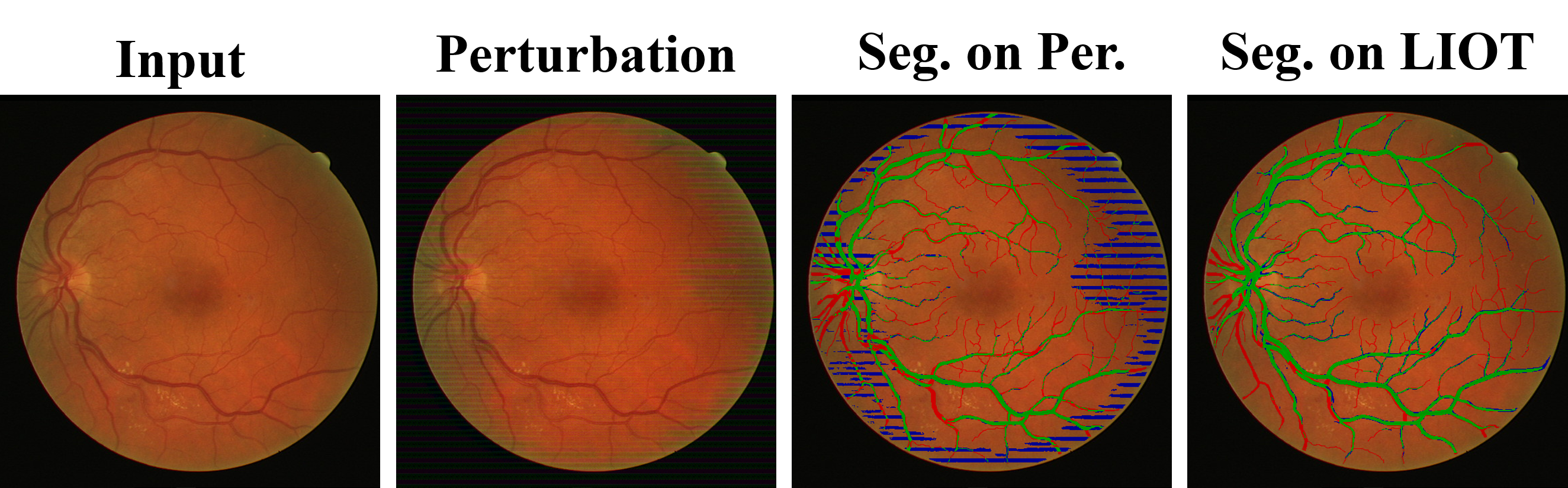}
\caption{In-dataset adversarial perturbation validation on DRIVE}
\end{subfigure}
\begin{subfigure}[b]{0.95\linewidth}
\centering
\includegraphics[width=1.0\linewidth]{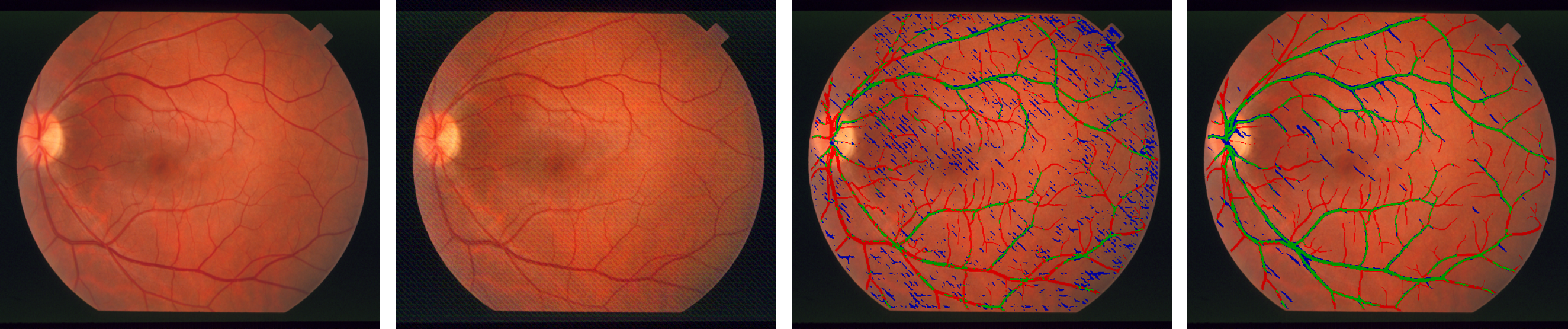}
\caption{In-dataset adversarial perturbation validation on STARE}
\end{subfigure}
\begin{subfigure}[b]{0.95\linewidth}
\centering
\includegraphics[width=1.0\linewidth]{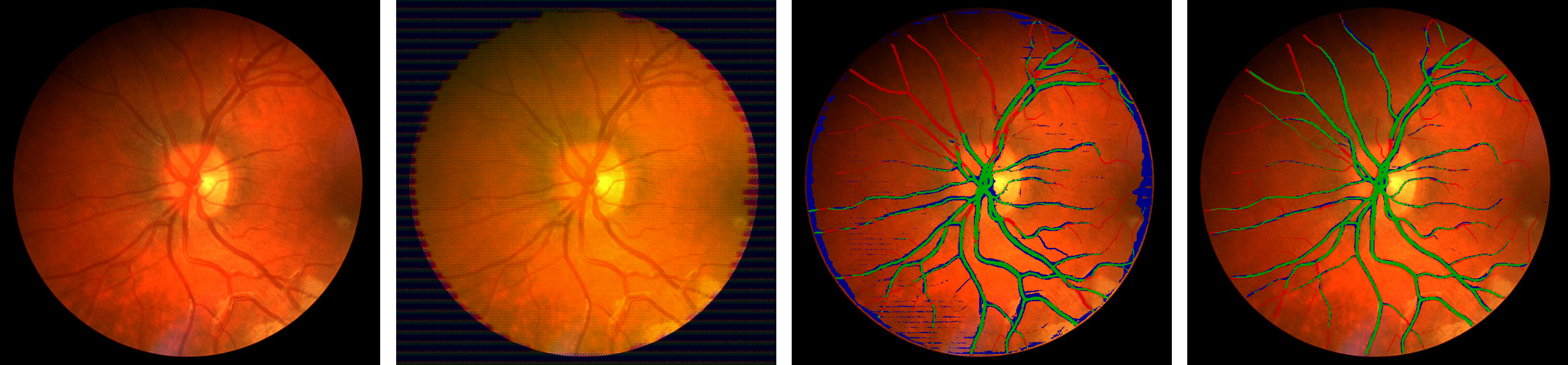}
\caption{In-dataset adversarial perturbation validation on CHASEDB1}
\end{subfigure}
\begin{subfigure}[b]{0.95\linewidth}
\centering
\includegraphics[width=1.0\linewidth]{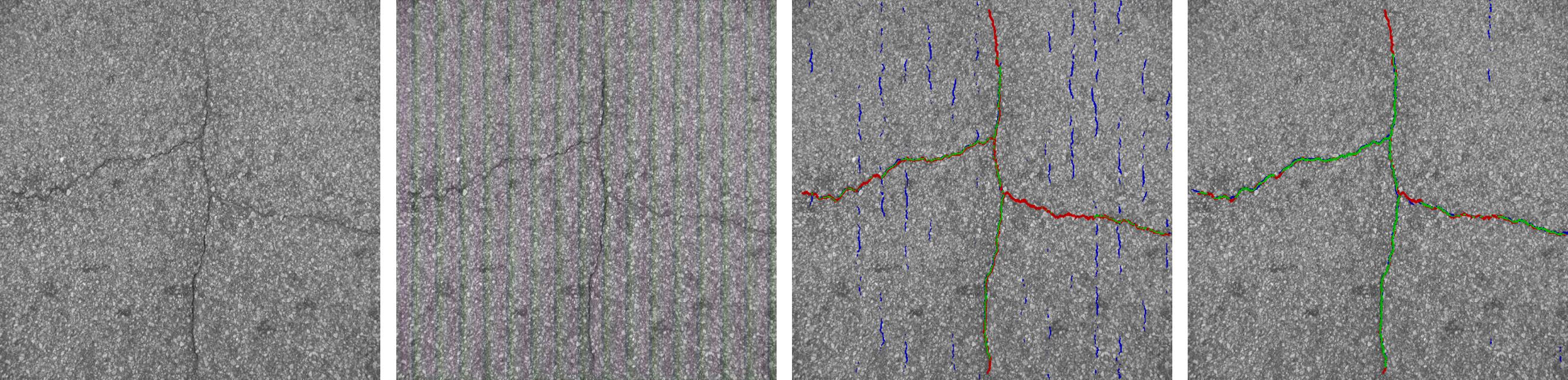}
\caption{In-dataset adversarial perturbation validation on Cracktree}
\end{subfigure}
\caption{Some segmentation results under in-dataset validation with non-targeted universal adversarial perturbation~\cite{poursaeed2018generative}. Green pixels: TPs; Red pixels: FNs; Blue pixels: FPs.  Best viewed by zooming in the electronic version.}
\label{fig:indatasetperturbation}
\end{figure}

\begin{figure*}[t]
\centering

\begin{subfigure}[b]{0.95\linewidth}
\centering
\includegraphics[width=1.0\linewidth]{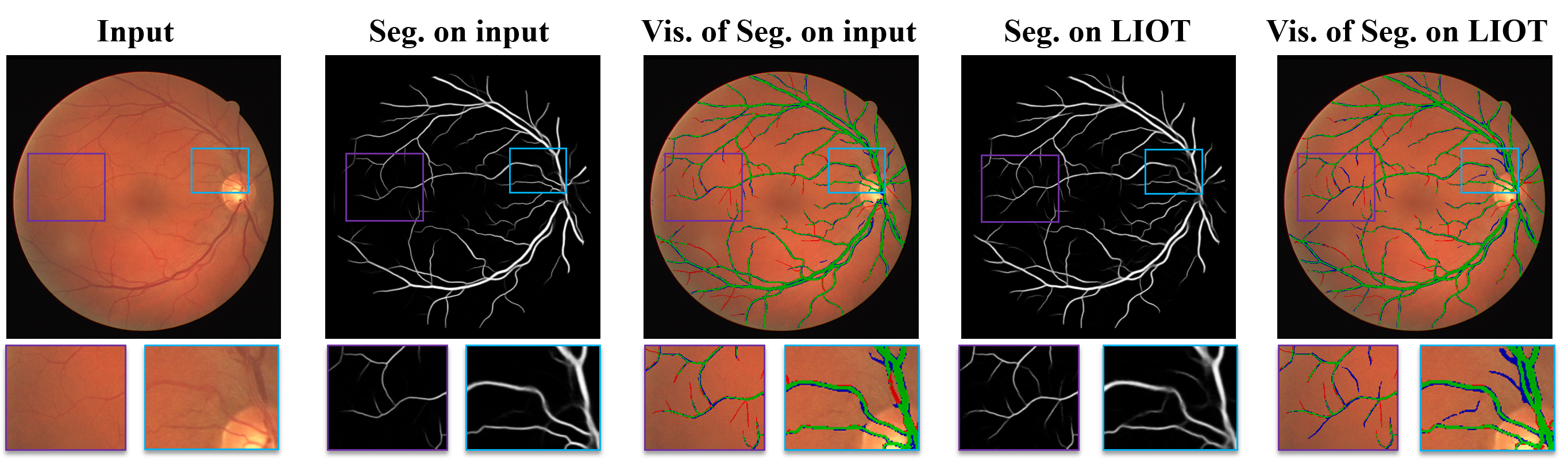}
\caption{Cross-dataset validation from CHASEDB1 to DRIVE}
\end{subfigure}
\begin{subfigure}[b]{0.95\linewidth}
\centering
\includegraphics[width=1.0\linewidth]{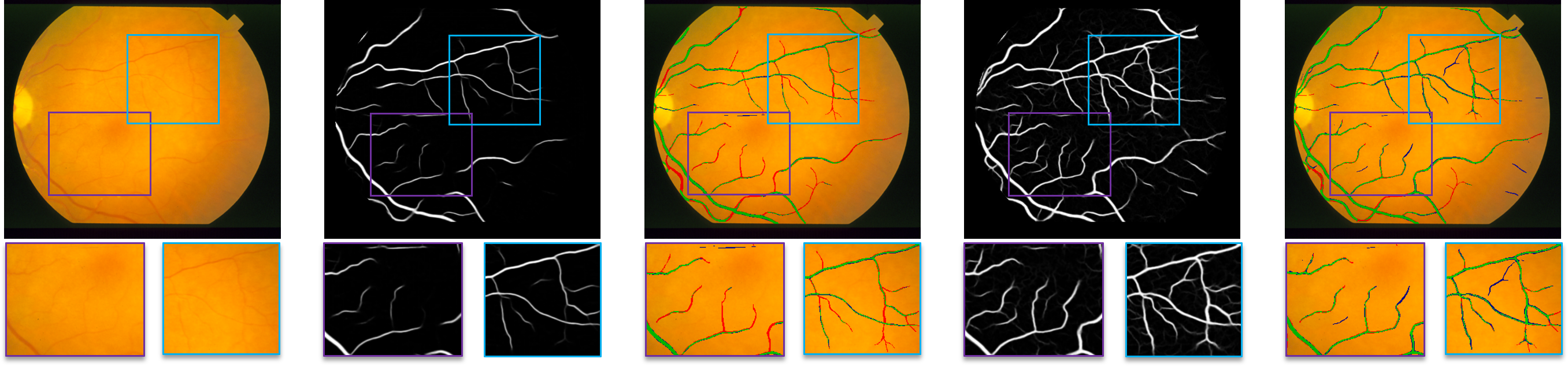}
\caption{Cross-dataset validation from DRIVE to STARE}
\end{subfigure}
\begin{subfigure}[b]{0.95\linewidth}
\centering
\includegraphics[width=1.0\linewidth]{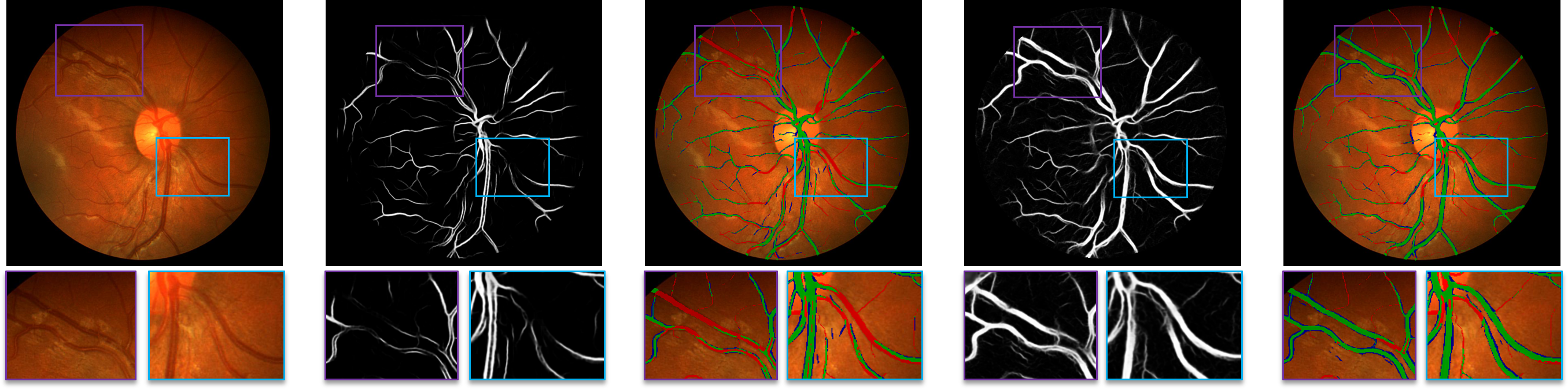}
\caption{Cross-dataset validation from STARE to CHASEDB1}
\end{subfigure}

\caption{Some segmentation results under cross-dataset validation. Green pixels: TPs; Red pixels: FNs; Blue pixels: FPs. Some FPs achieved by LIOT can find evidence in the original image, which might be TPs ignored in the manual annotation.  
}
\label{crossresult}
\end{figure*}

The proposed LIOT converts the original image into a relative order-based representation, which may lose some information on the image content. From this point of view, though the proposed LIOT mainly focus on improving the generalization performance for deep learning-based methods, we first conduct experiments on in-datasets: DRIVE, STARE, CHASEDB1, and CrackTree to show that the proposed LIOT does not significantly decrease the performance for the ability of model itself. 

Some qualitative results are illustrated in Fig.~\ref{fig:qualitative-indataset}. LIOT performs similarly with the baseline model, but usually has more false positives, which can find evidence in the original image, but are ignored in the ground-truth annotation. This is because that LIOT mainly focuses on the inherent characteristic (\eg, darker than the context) of curvilinear structure, and thus captures more objects having curvilinear structure. This may also explain why LIOT achieves slightly worse results than the baseline model on retinal images in terms of quantitative evaluation depicted in Table~\ref{indatasettab}. On the retinal datasets, LIOT is slightly worse than the baseline model, but performs much better than the variant of classical census transform. On the CrackTree, LIOT is very competitive with the baseline model. On average, LIOT is still comparable with the baseline model.

\subsection{In-dataset evaluation on images with perturbation}
\label{subsec:in-dataset-perturbation}
It is well known that deep learning-based models are highly vulnerable to adversarial perturbations, which may pose security problems for deep learning-based curvilinear object segmentation methods. In this section, we evaluate the robustness of the proposed LIOT against such adversarial perturbations. For that, we adopt the non-targeted universal perturbations by GAP~\cite{poursaeed2018generative} to generate adversarial perturbations. As shown in Fig.~\ref{fig:indatasetperturbation}, these perturbed images resemble the original images. We apply LIOT on the perturbed images and feed it to the trained model.  Some qualitative results are given in Fig.~\ref{fig:indatasetperturbation}. Though the perturbed images mislead both the trained baseline model and the model trained on LIOT, LIOT is more robust. 


The quantitative evaluation is depicted in Table~\ref{indatasetPerturbationtab}. The proposed LIOT outperforms the baseline model by a large margin (up to 47.8\%) in terms of all metrics. The proposed LIOT can effectively retain the inherent characteristic of curvilinear structure from the images with non-targeted universal adversarial perturbations, demonstrating the robustness of the proposed LIOT.




\setlength{\tabcolsep}{2.3pt}

\subsection{Cross retinal datasets evaluation}
\label{subsec:Cross-retinal}
We now evaluate the proposed LIOT by conducting cross-retinal dataset validation between the DRIVE, STARE, and CHASEDB1 dataset.


Some qualitative illustrations are shown in Fig.~\ref{crossresult}. As depicted in this figure, LIOT is able to correctly segment most retinal blood vessels, especially those of low contrast. The direct segmentation on the original image fails to retrieve those thin and low contrast retinal blood vessels. For instance, as depicted in Fig.~\ref{crossresult}(b), the model trained on DRIVE, using LIOT can accurately segment most retinal blood vessels, including the segments (blue pixels in the cropped region on the right side) of very low contrast and ignored even by the manual annotation. In fact, though these segments have very low contrast, we can still find cues in the original image and they are indeed retinal blood vessels. This demonstrates that LIOT is more robust to contrast changes. Indeed, as depicted in Fig.~\ref{fig:example}, the baseline method poorly segments the retinal blood vessels for the contrast changed image, on which LIOT still performs similarly with the original image. Thus, LIOT can improve the generalizability of some deep learning-based methods in segmenting retinal blood vessels.

The quantitative cross retinal dataset evaluation of LIOT is depicted in Table~\ref{retinaltable}. 
We randomly divide three times DRIVE, STARE, and CHASEDB datasets into train/test images, and we train the corresponding model three times. The mean and standard errors are reported.
Compared with the baseline method of IterNet~\cite{li2019iternet} which directly operates on the original image, LIOT outperforms or is on par with the baseline model in terms of classical pixel-level metrics: Se, Acc, AUC, and F1-score for all set of cross-dataset evaluation. LIOT has better mean values and smaller
standard errors in most cases. LIOT is also competitive with other state-of-the-art methods~\cite{li2015cross, oliveira2018retinal, yan2018joint, cherukuri2019deep, guo2019bts} dedicated for cross-retinal dataset evaluation. On average, LIOT performs much better than the baseline model and census transform based on all metrics except Sp. In particular, LIOT features a high sensitivity regime, which is important in clinical usage. Some ``false positives'' that are retinal blood vessels of low contrast and ignored by the manual annotation (described above) may degrade a bit the other classical metrics, such as the Sp. Therefore, LIOT may even perform slightly better in practice.

It is noteworthy that LIOT not only improves the pixel-level metrics, but also boosts the segmentation performance on the vessel network level. Specifically, as shown in Table~\ref{retinaltable}, LIOT consistently outperforms the baseline models in terms of connectivity metric. 
This implies that LIOT improves the integrity of the segmentation effect because connectivity assesses the fragmentation degree between the segmentation result and the ground truth. Besides, the connectivity is also an important clue for clinicians to calculate the complexity and density of branching of the retinal vascular tree to measure patient's condition~\cite{torp2018temporal}. Therefore, LIOT is also potentially useful for clinical disease diagnosis.

Though the goal is not to demonstrate that simply changing the input image by the proposed LIOT on a baseline method can outperform all other methods, as depicted in Table~\ref{retinaltable}, the experimental results show that the proposed method is competitive with all other methods~\cite{li2015cross, oliveira2018retinal, yan2018joint, cherukuri2019deep, guo2019bts} under cross-dataset validation, demonstrating the potential of the proposed LIOT in segmenting retinal blood vessels for a wide daily clinical usage.

\begin{table*}[htb]
\caption{Quantitative comparison of LIOT and some other methods under cross-dataset evaluation on retinal images. 
The results obtained from re-implementation are marked with $^{\dagger }$, and $*$ indicates the reported results from the original papers.}\label{retinaltable}
\centering
\begin{tabular}{|c|c|c|c|c|c|c|c|}
\hline
Cross-dataset                              & Methods             & Se    & Sp      & Acc      & AUC      & F1  &Connectivity     \\ \hline
\multirow{10}{1.7cm}{\centering STARE\\$\Rightarrow$\\DRIVE}   & ~\cite{li2015cross}$*$  & 0.727& 0.981   & 0.949   & 0.968   & -- & --       \\ 
& ~\cite{oliveira2018retinal}$*$        & 0.671& \textbf{0.992}  & 0.951   & 0.975   & -- & --       \\
                                          & ~\cite{yan2018joint}$*$           & 0.729& 0.982  & 0.949   & 0.960   & -- & --       \\
                                          & ~\cite{cherukuri2019deep}$*$   & \textbf{0.772}& 0.983   & \textbf{0.956}    & \textbf{0.977}   & -- & --       \\
                                          & ~\cite{guo2019bts}$*$        & 0.745& 0.978  & 0.950   & 0.971   & -- & --       \\ \cline{2-8}
                                          & Baseline~\cite{li2019iternet}$^{\dagger }$  & 0.781\textbf{$\pm$0.003}& \textbf{0.967$\pm$0.004}  & \textbf{0.943$\pm$0.003}   & 0.959$\pm$0.003   & \textbf{0.779$\pm$0.013} &0.779\textbf{$\pm$0.004}      \\
                                          & Census~\cite{zabih1996non}$^{\dagger }$  & 0.724$\pm$0.013& 0.955$\pm$0.006  & 0.926$\pm$0.007   & 0.932$\pm$0.009   & 0.716$\pm$0.022 &0.721$\pm$0.020      \\
                                          & LIOT$^{\dagger }$ & \textbf{0.791}$\pm$0.004& 0.963$\pm$0.006  & 0.941$\pm$0.004   & \textbf{0.962$\pm$0.002}   & 0.775$\pm$0.015 &\textbf{0.788}$\pm$0.009 \\
                                          \cline{2-8}
                                          & Topo~\cite{hu2019topology}$^{\dagger }$  & 0.770\textbf{$\pm$0.004}& \textbf{0.969$\pm$0.001} & \textbf{0.944$\pm$0.001}   & \textbf{0.954$\pm$0.002}   & \textbf{0.776}$\pm$0.006 &0.772\textbf{$\pm$0.004}\\
                                          & Topo+LIOT$^{\dagger }$  & \textbf{0.790}$\pm$0.006& 0.957$\pm$0.002 & 0.936\textbf{$\pm$0.001}   & 0.950$\pm$0.007   & 0.759\textbf{$\pm$0.001} &\textbf{0.792}$\pm$0.006\\\hline
\multirow{7}{1.7cm}{\centering CHASEDB1\\$\Rightarrow$\\DRIVE}& ~\cite{li2015cross}$*$& \textbf{0.731}& \textbf{0.981}  & \textbf{0.948}   & \textbf{0.961}   & -- & --       \\
                                          & ~\cite{guo2019bts}$*$                       & 0.696 & 0.970  & 0.938   & 0.952   & -- & -- \\ \cline{2-8}
                                          & Baseline~\cite{li2019iternet}$^{\dagger }$  & 0.763$\pm$0.030& \textbf{0.971$\pm$0.001}  & \textbf{0.945}$\pm$0.005   & 0.956$\pm$0.011   & \textbf{0.779}$\pm$0.023 &0.762$\pm$0.025       \\
                                          & Census~\cite{zabih1996non}$^{\dagger }$  & 0.719$\pm$0.024& 0.966$\pm$0.007  & 0.935\textbf{$\pm$0.003}   & 0.943$\pm$0.005   & 0.738\textbf{$\pm$0.007} &0.724$\pm$0.024      \\ 
                                          & LIOT$^{\dagger }$ & \textbf{0.799$\pm$0.007}& 0.963$\pm$0.005  & 0.942$\pm$0.005    & \textbf{0.966$\pm$0.003}   & \textbf{0.779}$\pm$0.019 &\textbf{0.798$\pm$0.003}\\
                                          \cline{2-8}
                                          & Topo~\cite{hu2019topology}$^{\dagger }$  & 0.750$\pm$0.036& \textbf{0.968}$\pm$0.002 & 0.941$\pm$0.006   & 0.933$\pm$0.026   & 0.762$\pm$0.027 &0.753$\pm$0.036\\
                                          & Topo+LIOT$^{\dagger }$  & \textbf{0.781$\pm$0.010}& 0.965\textbf{$\pm$0.001} & \textbf{0.942$\pm$0.001}   & \textbf{0.960$\pm$0.005}   & \textbf{0.773$\pm$0.006} &\textbf{0.784$\pm$0.010}\\  \hline \hline
\multirow{10}{1.7cm}{\centering DRIVE\\$\Rightarrow$\\STARE}& ~\cite{li2015cross}$*$   & 0.703& 0.983  & 0.955   & 0.967   & -- & --      \\
                                          & ~\cite{oliveira2018retinal}$*$ & \textbf{0.845}& 0.973  & 0.960   & \textbf{0.985}   & -- & --      \\
                                          & ~\cite{yan2018joint}$*$             & 0.721& 0.984   & 0.957   & 0.971   & -- & --      \\
                                          & ~\cite{cherukuri2019deep}$*$     & 0.778& \textbf{0.986}  & \textbf{0.971}   & 0.982   & -- & --     \\
                                          & ~\cite{guo2019bts}$*$        & 0.719& 0.982  & 0.955   & 0.969   & -- & --      \\ \cline{2-8}
                                          & Baseline~\cite{li2019iternet}$^{\dagger }$  & 0.760$\pm$0.038 & \textbf{0.978}$\pm$0.002  & 0.956$\pm$0.003  & 0.962$\pm$0.010   & 0.782$\pm$0.018 & 0.731$\pm$0.052      \\
                                          & Census~\cite{zabih1996non}$^{\dagger }$  & 0.766$\pm$0.011 & 0.970$\pm$0.005  & 0.949$\pm$0.005   & 0.966$\pm$0.003   & 0.759$\pm$0.020 & 0.720$\pm$0.058      \\
                                          & LIOT$^{\dagger }$ & \textbf{0.812$\pm$0.007} &0.975\textbf{$\pm$0.001} &\textbf{0.958$\pm$0.001} &\textbf{0.981$\pm$0.002} &\textbf{0.800$\pm$0.006} &\textbf{0.806$\pm$0.010}   \\
                                          \cline{2-8}
                                          & Topo~\cite{hu2019topology}$^{\dagger }$  & 0.713$\pm$0.031 & \textbf{0.977}$\pm$0.002 &0.949$\pm$0.004   & 0.915$\pm$0.034  & 0.745$\pm$0.023 &0.673$\pm$0.046\\
                                          & Topo+LIOT$^{\dagger }$  & \textbf{0.799$\pm$0.005}& 0.972\textbf{$\pm$0.001} & \textbf{0.955$\pm$0.001}   & \textbf{0.977$\pm$0.003}   & \textbf{0.785$\pm$0.006} &\textbf{0.796$\pm$0.006}\\ \hline
\multirow{7}{1.7cm}{\centering CHASEDB1\\ $\Rightarrow$ \\STARE} & ~\cite{li2015cross}$*$   & \textbf{0.694}& \textbf{0.983}  & \textbf{0.954}   & \textbf{0.962}    & -- & -- \\
                                          & ~\cite{guo2019bts}$*$    & 0.680& 0.981  & 0.950   & 0.952   & --  & --\\ \cline{2-8}
                                          & Baseline~\cite{li2019iternet}$^{\dagger }$  & 0.722$\pm$0.020& \textbf{0.976}$\pm$0.002  & 0.950\textbf{$\pm$0.001}   & 0.942$\pm$0.014   & 0.750$\pm$0.011 &0.704$\pm$0.018        \\
                                          & Census~\cite{zabih1996non}$^{\dagger }$  & 0.764$\pm$0.009& 0.968$\pm$0.002  & 0.947\textbf{$\pm$0.001}   & 0.964$\pm$0.004   & 0.751\textbf{$\pm$0.003} &0.727$\pm$0.061      \\
                                          & LIOT$^{\dagger }$ & \textbf{0.807$\pm$0.008} & 0.971\textbf{$\pm$0.001}  & \textbf{0.954$\pm$0.001}   & \textbf{0.978$\pm$0.001} & \textbf{0.784}$\pm$0.007 &\textbf{0.804$\pm$0.004} \\
                                          \cline{2-8}
                                          & Topo~\cite{hu2019topology}$^{\dagger }$  & 0.727$\pm$0.026& \textbf{0.974$\pm$0.002} & 0.948$\pm$0.004   & 0.917$\pm$0.030   & 0.743$\pm$0.022 &0.710$\pm$0.026\\
                                          & Topo+LIOT$^{\dagger }$  & \textbf{0.796$\pm$0.016}& \textbf{0.974$\pm$0.002} & \textbf{0.956$\pm$0.001}   & \textbf{0.976$\pm$0.003}   & \textbf{0.788$\pm$0.020} &\textbf{0.795$\pm$0.015}\\  \hline \hline
\multirow{7}{1.7cm}{\centering DRIVE\\\ $\Rightarrow$ \\CHASEDB1} & ~\cite{li2015cross}$*$   & \textbf{0.712}& \textbf{0.979}  & 0.943   & \textbf{0.963}   & -- & --       \\
                                          & ~\cite{guo2019bts}$*$        & 0.698 & 0.972  & \textbf{0.944}   & 0.957   & -- & --       \\ \cline{2-8}
                                          & Baseline~\cite{li2019iternet}$^{\dagger }$  & 0.702$\pm$0.036 & 0.963$\pm$0.005  & 0.939$\pm$0.007   & 0.937$\pm$0.013   & 0.682$\pm$0.044 & 0.697$\pm$0.041        \\
                                          & Census~\cite{zabih1996non}$^{\dagger }$ & 0.682\textbf{$\pm$0.005}& 0.958$\pm$0.003  & 0.933$\pm$0.003   & 0.939$\pm$0.004   & 0.650\textbf{$\pm$0.012} &0.634$\pm$0.071     \\
                                          & LIOT$^{\dagger }$ & \textbf{0.784}$\pm$0.011 &\textbf{0.968$\pm$0.002}  &\textbf{0.951$\pm$0.002}  &\textbf{0.973$\pm$0.002}  &\textbf{0.749}$\pm$0.020 &\textbf{0.780$\pm$0.005}\\ 
                                          \cline{2-8}
                                          & Topo~\cite{hu2019topology}$^{\dagger }$  & 0.682$\pm$0.077 & 0.957$\pm$0.005 & 0.932$\pm$0.010   & 0.912$\pm$0.042   & 0.649$\pm$0.065 & 0.676$\pm$0.084\\
                                          & Topo+LIOT$^{\dagger }$  & \textbf{0.769$\pm$0.011}& \textbf{0.962$\pm$0.002} & \textbf{0.944$\pm$0.002}   & \textbf{0.966$\pm$0.001}   & \textbf{0.720$\pm$0.002} &\textbf{0.772$\pm$0.012}\\\hline
\multirow{7}{1.7cm}{\centering STARE\\ $\Rightarrow$ \\CHASEDB1}   & ~\cite{li2015cross}$*$    & \textbf{0.724}  & \textbf{0.977}  & \textbf{0.942}  & \textbf{0.955}    & --  & --       \\
                                          & ~\cite{guo2019bts}$*$        & 0.673 & 0.971   & 0.941   & 0.951   & --   & --      \\ \cline{2-8}
                                          & Baseline~\cite{li2019iternet}$^{\dagger }$  & 0.653$\pm$0.101& 0.956\textbf{$\pm$0.003}  & 0.926$\pm$0.012   & 0.908$\pm$0.048   & 0.635$\pm$0.075 &0.654$\pm$0.097         \\
                                          & Census~\cite{zabih1996non}$^{\dagger }$  & 0.595$\pm$0.049& 0.951$\pm$0.004  & 0.918$\pm$0.006   & 0.891$\pm$0.023   & 0.568$\pm$0.036 &0.552$\pm$0.094      \\ 
                                          & LIOT$^{\dagger }$ & \textbf{0.764$\pm$0.008} & \textbf{0.959$\pm$0.003}   & \textbf{0.939$\pm$0.004}   & \textbf{0.961$\pm$0.003}    & \textbf{0.717$\pm$0.016} &\textbf{0.760$\pm$0.007}\\\cline{2-8}
                                          & Topo~\cite{hu2019topology}$^{\dagger }$  & 0.630$\pm$0.106& \textbf{0.961}$\pm$0.004 & 0.929$\pm$0.010   & 0.872$\pm$0.078   & 0.633$\pm$0.073 &0.631$\pm$0.103\\
                                          & Topo+LIOT$^{\dagger }$  & \textbf{0.759$\pm$0.011}& 0.960\textbf{$\pm$0.003} & \textbf{0.940$\pm$0.002}   & \textbf{0.954$\pm$0.006}   & \textbf{0.716$\pm$0.003} &\textbf{0.761$\pm$0.013}\\                                          \hline \hline
\multirow{5}{1.7cm}{\centering Average} 
                                          & Baseline~\cite{li2019iternet}$^{\dagger }$  & 0.730$\pm$0.038& \textbf{0.969$\pm$0.003}  & 0.943$\pm$0.005   & 0.944$\pm$0.017   & 0.735$\pm$0.031 & 0.721$\pm$0.040        \\
                                          & Census~\cite{zabih1996non}$^{\dagger }$  & 0.708$\pm$0.019 & 0.962$\pm$0.005   & 0.935$\pm$0.004   &0.939$\pm$0.008    &0.697$\pm$0.017 & 0.680$\pm$0.055       \\ 
                                          & LIOT$^{\dagger }$ & \textbf{0.793$\pm$0.008} & 0.966\textbf{$\pm$0.003}   & \textbf{0.947$\pm$0.003}   & \textbf{0.970$\pm$0.002}   & \textbf{0.767$\pm$0.014} &\textbf{0.789$\pm$0.006}\\  \cline{2-8}
                                          & Topo~\cite{hu2019topology}$^{\dagger }$  & 0.712$\pm$0.047& \textbf{0.968}$\pm$0.003 & 0.940$\pm$0.006   & 0.917$\pm$0.035   & 0.718$\pm$0.036 & 0.702$\pm$0.050\\
                                          & Topo+LIOT$^{\dagger }$  & \textbf{0.782$\pm$0.010}& 0.965\textbf{$\pm$0.002} & \textbf{0.945$\pm$0.001}   &\textbf{0.964$\pm$0.004}   & \textbf{0.757$\pm$0.004} &\textbf{0.783$\pm$0.010}\\ 
\hline
\end{tabular}
\end{table*}

\subsection{Cross CrackTree and retinal datasets evaluation}
\label{subsec:Cross-crack}
\setlength{\tabcolsep}{2.3pt}
\begin{table}[ht]
\caption{Cross retinal and crack dataset evaluation of LIOT.}\label{cracktab}
\centering
\begin{tabular}{|c|c|c|c|c|c|c|c|}
\hline
Cross-dataset                              & Methods             & Se    & Sp      & Acc      & AUC      & F1    &Connectivity  \\ \hline
\multirow{5}{1.5cm}{\centering CrackTree\\$\Rightarrow$\\ DRIVE}
                                          & Baseline~\cite{li2019iternet}  & 0.131& \textbf{0.965}  & 0.859   & 0.646   & 0.191 &0.008\\
                                          & Census~\cite{zabih1996non}  & 0.532& 0.896  & 0.850   & 0.812   & 0.474 &0.487      \\
                                          & LIOT & \textbf{0.712} &0.895  &\textbf{0.872}   &\textbf{0.899}   &\textbf{0.585} &\textbf{0.675}\\\cline{2-8}
                                          & Topo~\cite{hu2019topology}  & 0.013& \textbf{0.974} & 0.852   & 0.504   & 0.023 &0.014\\
                                          & Topo+LIOT  & \textbf{0.655}& 0.902 & \textbf{0.871}   & \textbf{0.860}   & \textbf{0.563} &\textbf{0.656}\\\hline
\multirow{5}{1.5cm}{\centering CrackTree \\$\Rightarrow$\\ STARE}  
                                          & Baseline~\cite{li2019iternet}  & 0.130& \textbf{0.969}  & 0.882   & 0.626   & 0.185 &0.006\\
                                          & Census~\cite{zabih1996non}  & 0.436& 0.892  & 0.845   & 0.750   & 0.366      &0.239\\
                                          & LIOT & \textbf{0.640} & 0.911   & \textbf{0.883}   & \textbf{0.898}    & \textbf{0.530} &\textbf{0.505}\\\cline{2-8}
                                          & Topo~\cite{hu2019topology}  & 0.004& \textbf{0.991} & \textbf{0.889}   & 0.489   & 0.007 &0.004\\
                                          & Topo+LIOT  & \textbf{0.539}& 0.928 & 0.888   & \textbf{0.814}   & \textbf{0.497} &\textbf{0.541}\\\hline
\multirow{5}{1.5cm}{\centering CrackTree \\$\Rightarrow$\\ CHASEDB1}  
                                          & Baseline~\cite{li2019iternet}  & 0.044& \textbf{0.981}  & \textbf{0.897}   & 0.573   & 0.071 &0.010\\
                                          & Census~\cite{zabih1996non}  & 0.369& 0.922  & 0.872   & 0.753   & 0.342      &0.334\\
                                          & LIOT & \textbf{0.650} & 0.890   & 0.877   & \textbf{0.895}    & \textbf{0.489} &\textbf{0.580}\\\cline{2-8}
                                          & Topo~\cite{hu2019topology}  & 0.003& \textbf{0.990} & \textbf{0.892}   & 0.486   & 0.005 &0.003\\
                                          & Topo+LIOT  & \textbf{0.623}& 0.908 & 0.879   & \textbf{0.853}   & \textbf{0.507} &\textbf{0.617}\\\hline \hline   
\multirow{5}{1.5cm}{\centering DRIVE\\$\Rightarrow$\\CrackTree}   & Baseline~\cite{li2019iternet}  & 0.037& \textbf{0.997}  & 0.984   & 0.643   & 0.061   &0.015   \\
                                            & Census~\cite{zabih1996non}  & 0.349& 0.995  & 0.985   & 0.828   & 0.408  &0.409    \\
                                            & LIOT & \textbf{0.485}& 0.994  & \textbf{0.987}   & \textbf{0.911}   & \textbf{0.515} &\textbf{0.527}\\ \cline{2-8}
                                          & Topo~\cite{hu2019topology}  & 0.212& 0.752 & 0.744   & 0.425   & 0.023 &0.229\\
                                          & Topo+LIOT  &  \textbf{0.389}&  \textbf{0.995} &  \textbf{0.986}   &  \textbf{0.905}   &  \textbf{0.444} & \textbf{0.383}\\ \hline
\multirow{5}{1.5cm}{\centering STARE\\$\Rightarrow$\\CrackTree} & Baseline~\cite{li2019iternet}  & 0.022& 0.988  & 0.974   & 0.371   & 0.024  &0.004     \\
                                        & Census~\cite{zabih1996non}  & 0.040& \textbf{0.999}  & \textbf{0.986}   & 0.696   & 0.075      &0.103\\
                                        & LIOT & \textbf{0.290}& 0.995  & 0.985    & \textbf{0.865}   & \textbf{0.360}  &\textbf{0.400}\\\cline{2-8}
                                          & Topo~\cite{hu2019topology}  & 0.042& 0.970 & 0.957   & 0.505   & 0.027 &0.040\\
                                          & Topo+LIOT  &  \textbf{0.209}&  \textbf{0.997} &  \textbf{0.985}   &  \textbf{0.784}   &  \textbf{0.289 }& \textbf{0.207}\\ \hline
\multirow{5}{1.5cm}{\centering CHASEDB1 \\$\Rightarrow$\\ CrackTree}   
                                          & Baseline~\cite{li2019iternet}  & 0.099  & 0.994   & 0.981   & 0.684 &0.129   &0.028   \\
                                          & Census~\cite{zabih1996non}  & 0.178&\textbf{0.998}  & 0.986   & 0.726   & 0.272  &0.254    \\
                                          & LIOT & \textbf{0.452} &0.996  &\textbf{0.988}  &\textbf{0.909}  &\textbf{0.516}   &\textbf{0.518}\\\cline{2-8}
                                          & Topo~\cite{hu2019topology}  & 0.150& 0.977 & 0.965   & 0.733   & 0.110 &0.147\\
                                          & Topo+LIOT  & \textbf{0.454}& \textbf{0.994} & \textbf{0.987}   & \textbf{0.888}   & \textbf{0.493} &\textbf{0.456}\\ \hline \hline
\multirow{5}{1.5cm}{\centering Average}  
                                          & Baseline~\cite{li2019iternet}  &0.077 & \textbf{0.982}  &0.930    &0.591    &0.110 &0.012\\
                                          & Census~\cite{zabih1996non}  &0.317 &0.950   &0.921    &0.761    &0.323       &0.304\\
                                          & LIOT & \textbf{0.538} &0.947    & \textbf{0.932}   & \textbf{0.896}    & \textbf{0.499} &\textbf{0.534}\\\cline{2-8}
                                          & Topo~\cite{hu2019topology}  & 0.160& 0.942 & 0.883   & 0.524   & 0.033 &0.073\\  
                                          & Topo+LIOT  & \textbf{0.479}& \textbf{0.954} & \textbf{0.933}   & \textbf{0.851}   & \textbf{0.466} &\textbf{0.477}\\        
\hline
\end{tabular}
\end{table}

In addition to cross retinal dataset (with small domain gaps) validation depicted in Section~\ref{subsec:Cross-retinal}, We also evaluate the proposed LIOT by conducting cross-dataset validation between the retinal datasets and CrackTree dataset. To the best of our knowledge, no previous research has investigated capturing the inherent characteristic of curvilinear structure for such large appearance gaps. 

\begin{figure}[ht]
\centering

\begin{subfigure}[b]{0.95\linewidth}
\centering
\includegraphics[width=1.0\linewidth]{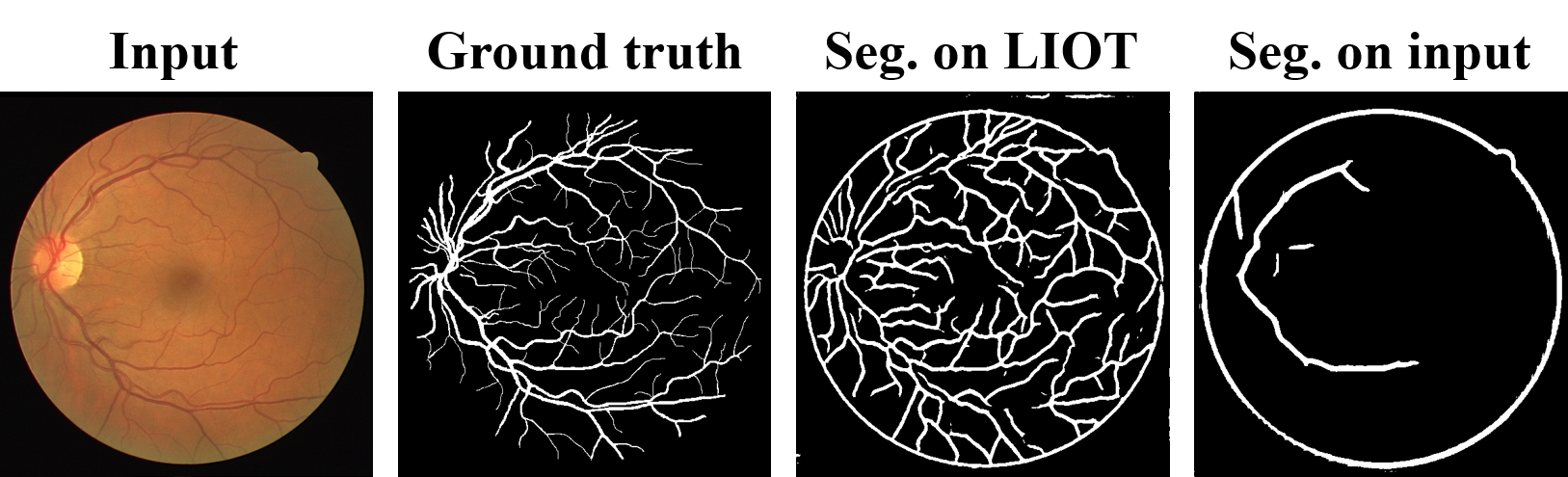}
\caption{Cross-dataset validation from CrackTree to DRIVE}
\end{subfigure}
\begin{subfigure}[b]{0.95\linewidth}
\centering
\includegraphics[width=1.0\linewidth]{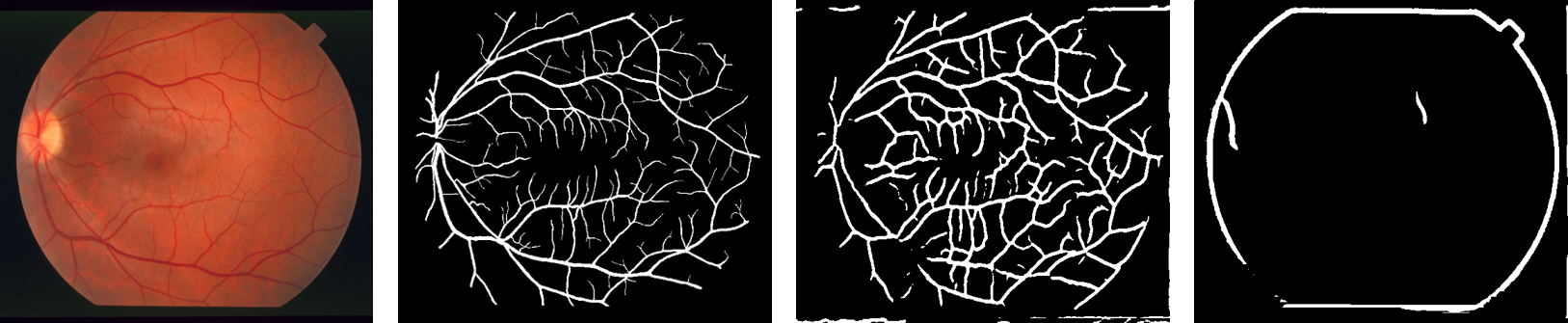}
\caption{Cross-dataset validation from CrackTree to STARE}
\end{subfigure}
\begin{subfigure}[b]{0.95\linewidth}
\centering
\includegraphics[width=1.0\linewidth]{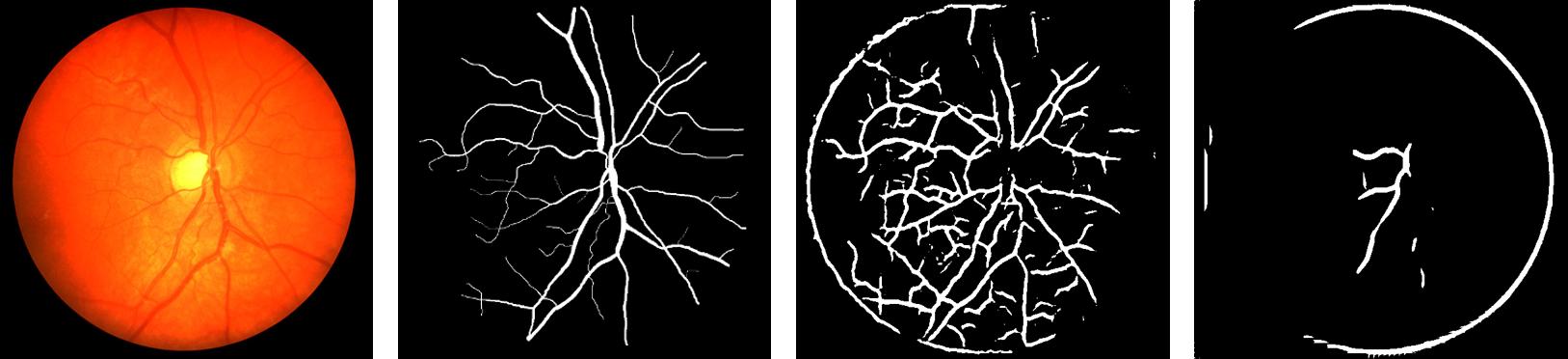}
\caption{Cross-dataset validation from CrackTree to CHASEDB1}
\end{subfigure}

\caption{Visualization of the segmentation results under cross-dataset validation between CrackTree and retinal dataset.}
\label{crossretinal_crackresult}
\end{figure}

\begin{figure}[t]
\centering

\begin{subfigure}[b]{0.95\linewidth}
\centering
\includegraphics[width=1.0\linewidth]{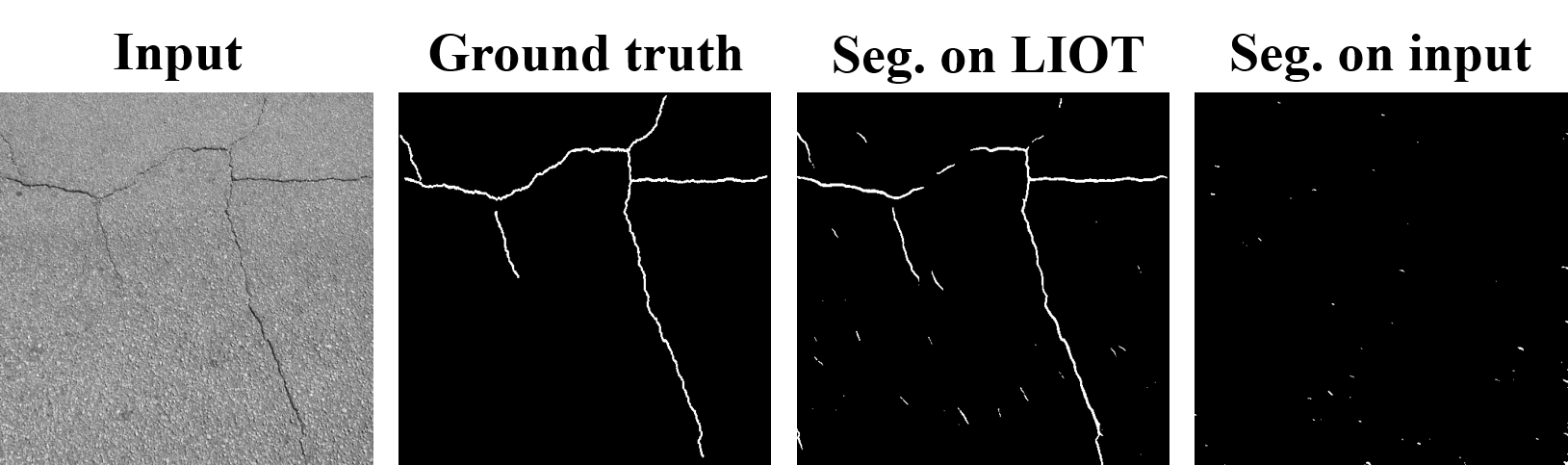}
\caption{Cross-dataset validation from DRIVE to CrackTree}
\end{subfigure}
\begin{subfigure}[b]{0.95\linewidth}
\centering
\includegraphics[width=1.0\linewidth]{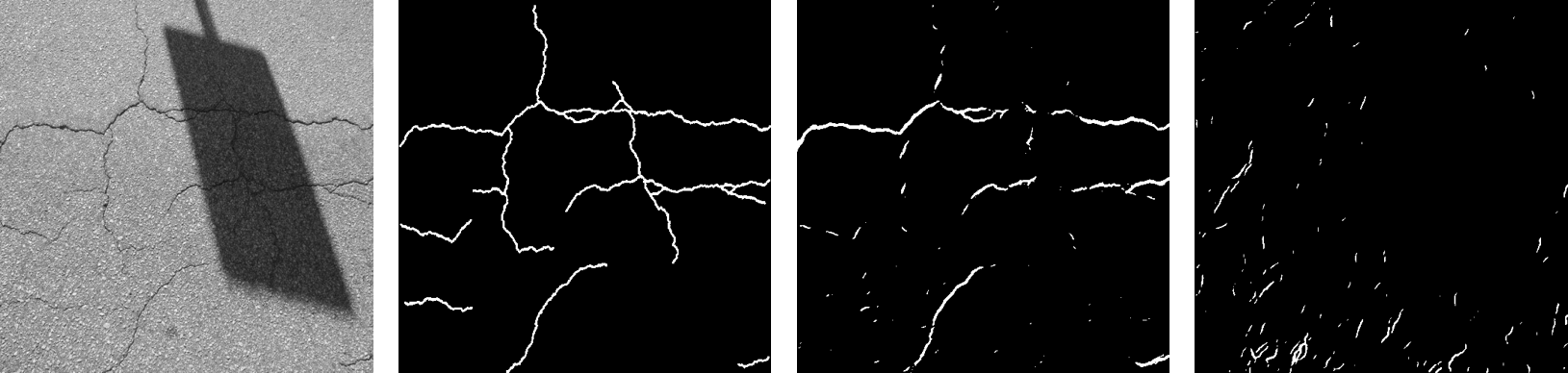}
\caption{Cross-dataset validation from STARE to CrackTree}
\end{subfigure}
\begin{subfigure}[b]{0.95\linewidth}
\centering
\includegraphics[width=1.0\linewidth]{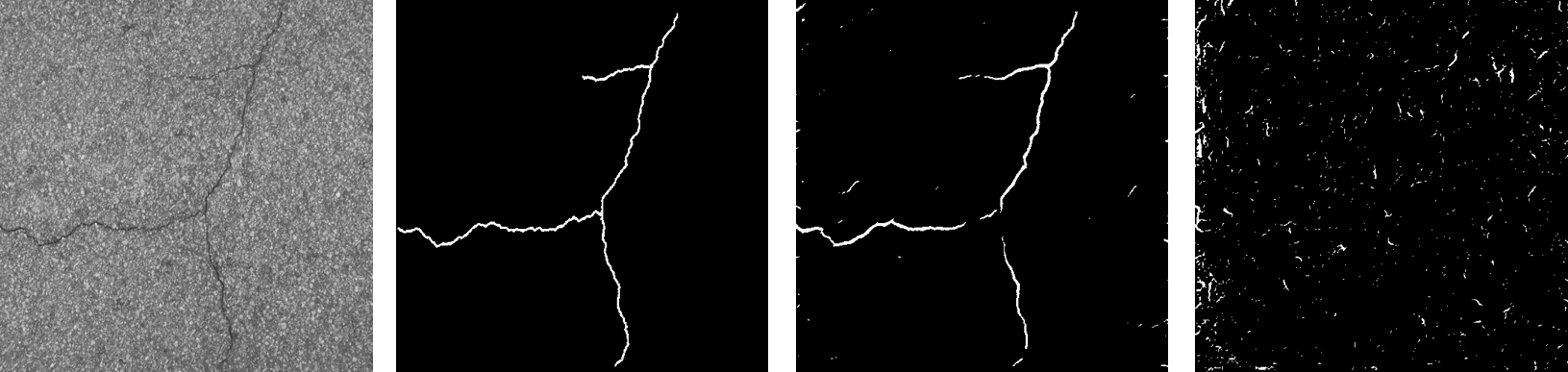}
\caption{Cross-dataset validation from CHASEDB1 to CrackTree}
\end{subfigure}

\caption{Visualization of the segmentation results under cross-dataset validation between retinal and CrackTree datasets.}
\label{crossCrack_retinalresult}
\end{figure}

\medskip
\noindent
\textbf{Cross-dataset validation from CrackTree to retinal datasets:}
As shown in Fig.~\ref{crossCrack_retinalresult}, the CrackTree dataset has a great appearance gap with retinal images. We first apply the model trained on the CrackTree dataset to segment retinal blood vessels. Some qualitative retinal blood vessel segmentation results using the model trained on the CrackTree dataset are given in Fig.~\ref{crossretinal_crackresult}. The proposed LIOT roughly segment the curvilinear structure in retinal images. From this figure, we can observe that LIOT preserves the inherent characteristic of curvilinear structure in retinal images. LIOT is also able to segment the curvilinear structure from various contexts. On the contrary, the baseline method can not capture the curvilinear structure of such large appearance gaps. The direct segmentation on the original image fails to retrieve most retinal blood vessels. These are common results as most deep-learning methods struggle for large image appearance gaps.

\begin{figure}[ht]
\centering

\begin{subfigure}[b]{0.95\linewidth}
\centering
\includegraphics[width=1.0\linewidth]{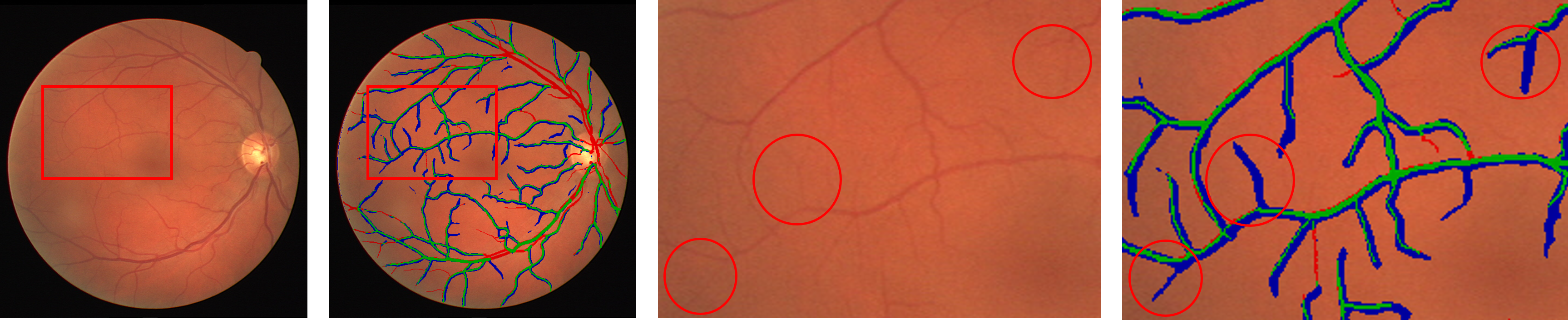}
\caption{Cross-dataset validation from CrackTree to DRIVE}
\end{subfigure}
\begin{subfigure}[b]{0.95\linewidth}
\centering
\includegraphics[width=1.0\linewidth]{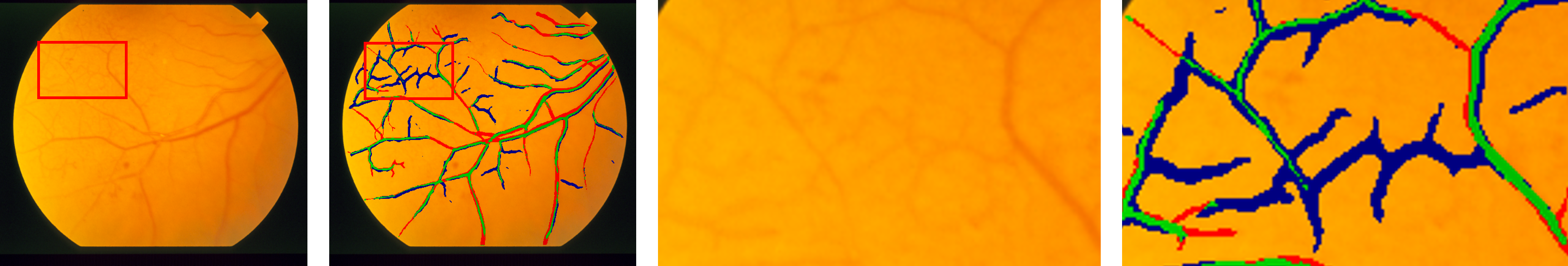}
\caption{Cross-dataset validation from CrackTree to STARE}
\end{subfigure}
\begin{subfigure}[b]{0.95\linewidth}
\centering
\includegraphics[width=1.0\linewidth]{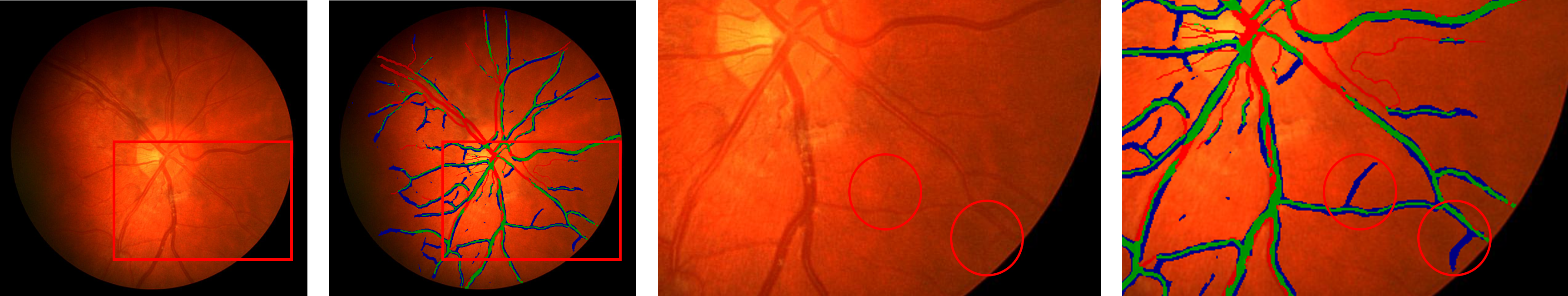}
\caption{Cross-dataset validation from CrackTree to CHASEDB1}
\end{subfigure}

\caption{Visualization of some segmentation detail under cross-dataset validation between CrackTree and retinal datasets. Green pixels: TPs; Red pixels: FNs; Blue pixels: FPs. Some FPs achieved by LIOT can find evidence in the original image, which might be TPs ignored in the manual annotation.}
\label{crosscrackdetail}
\end{figure}

More precisely, as depicted in Fig.~\ref{crosscrackdetail}, for the model trained on CrackTree, using LIOT can accurately capture most retinal blood vessels, including those blue pixels which are of very low contrast and ignored by the manual annotation. Such an effect is difficult to achieve by general methods, especially in a simple way like LIOT. This demonstrates that LIOT is robust to contrast changes and is able to capture the inherent curvilinear structure. 

The quantitative evaluation is shown in Table.~\ref{cracktab}. Compared with the baseline method, our proposed LIOT outperforms them by a large margin in terms of Se, AUC, and F1-score. Specifically, the proposed LIOT improves the baseline method by 34.5\% to 54.0\% in F1-score, significantly outperforming the baseline method. From the quantitative results given in Table~\ref{GaptoRtab}, the proposed LIOT makes the gap between CrackTree and the retinal dataset much smaller than the baseline method.
Thus, LIOT significantly improves the generalization performance of the baseline method, especially for the dataset that has curvilinear structures with great appearance gaps.

\medskip
\noindent
\textbf{Cross-dataset validation from retinal to CrackTree dataset:}
We also evaluate the proposed LIOT on CrackTree using the model trained on the retinal datasets. The CrackTree dataset has a great appearance gap compared with the retinal datasets, and also has multiple shadows in the context. Though this makes such cross-dataset segmentation a challenging task, the proposed LIOT still successfully segments most curvilinear structures in the CrackTree dataset. Some qualitative results are illustrated in Fig.~\ref{crossCrack_retinalresult}. As shown in Fig.~\ref{crossCrack_retinalresult}(b), LIOT is also able to extract curvilinear structure with the shadow influence. The quantitative compassion with the baseline method under this cross-dataset setting is given in Table.~\ref{cracktab}. LIOT achieves very competitive performance with the baseline method. Specifically, compared with the baseline model, LIOT achieves significant improvements, ranging from 26.2\% to 45.4\% in terms of F1-score. From the quantitative result given in Table~\ref{GaptoRtab}, the proposed LIOT makes the gap between retinal and crack images much smaller than the baseline method.

\medskip

From the qualitative results shown in Fig.~\ref{crossretinal_crackresult} and Fig.~\ref{crossCrack_retinalresult}, and quantitative evaluations depicted in Table.~\ref{cracktab} and Table.~\ref{GaptoRtab}, the proposed LIOT is capable to capture the curvilinear structure in various curvilinear object datasets. This demonstrates the generalization ability of the proposed LIOT. Simply changing the original input images with LIOT can help deep learning-based methods express a better generalization in various curvilinear object segmentation. 

\setlength{\tabcolsep}{4pt}
\renewcommand{\arraystretch}{1.1}
\begin{table}[t]
\caption{Quantitative F1-score gap between in-dataset (Tab.~\ref{indatasettab}) and cross-dataset results (Tab.~\ref{cracktab}).}\label{GaptoRtab}
\centering
\begin{tabular}{|c|c|c|c|c|}
\hline
Cross-dataset       & Baseline     & LIOT    & Topo  & Topo+LIOT \\ \hline
\multirow{1}{3.2cm}{\centering CrackTree$\Rightarrow$ DRIVE}   & 0.563 & \textbf{0.205}  & 0.746 & \textbf{0.238}\\ 
\multirow{1}{3.2cm}{\centering CrackTree$\Rightarrow$ STARE} & 0.569 &\textbf{0.260} & 0.762 &\textbf{0.304}\\ 
\multirow{1}{3.2cm}{\centering CrackTree $\Rightarrow$ CHASEDB1}  & 0.683  &\textbf{0.301} & 0.764 &\textbf{0.294}\\ 
\multirow{1}{3.2cm}{\centering DRIVE$\Rightarrow$ CrackTree}   & 0.760 & \textbf{0.299}  & 0.799 & \textbf{0.371}\\ 
\multirow{1}{3.2cm}{\centering STARE$\Rightarrow$ CrackTree} & 0.802 &\textbf{0.450} & 0.802 &\textbf{0.518}\\ 
\multirow{1}{3.2cm}{\centering CHASEDB1 $\Rightarrow$ CrackTree}  & 0.682 &\textbf{0.276} & 0.702 &\textbf{0.306}\\ \hline
\end{tabular}
\end{table}

\subsection{Generalization to images with different curvilinear objects}
\label{subsec:moreillustrations}

\begin{figure}[t]
\centering

\begin{subfigure}[b]{0.3\linewidth}
\centering
\includegraphics[width=1.0\linewidth]{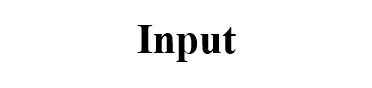}
\end{subfigure}
\hskip 0.1cm
\begin{subfigure}[b]{0.3\linewidth}
\centering
\includegraphics[width=1.0\linewidth]{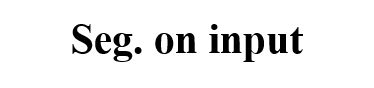}
\end{subfigure}
\hskip 0.1cm
\begin{subfigure}[b]{0.3\linewidth}
\centering
\includegraphics[width=1.0\linewidth]{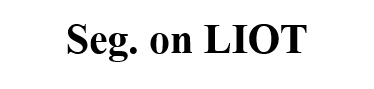}
\end{subfigure}

\vskip 0.001cm
\begin{subfigure}[b]{0.3\linewidth}
\centering
\includegraphics[width=1.0\linewidth]{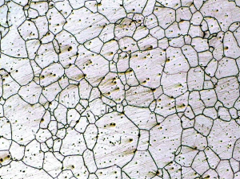}
\end{subfigure}
\hskip 0.1cm
\begin{subfigure}[b]{0.3\linewidth}
\centering
\includegraphics[width=1.0\linewidth]{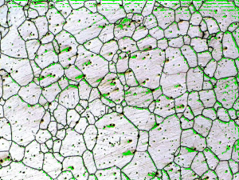}
\end{subfigure}
\hskip 0.1cm
\begin{subfigure}[b]{0.3\linewidth}
\centering
\includegraphics[width=1.0\linewidth]{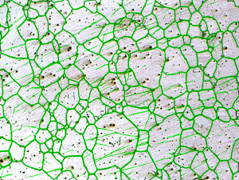}
\end{subfigure}

\vskip 0.1cm
\begin{subfigure}[b]{0.3\linewidth}
\centering
\includegraphics[width=1.0\linewidth]{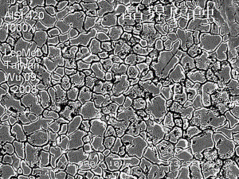}
\end{subfigure}
\hskip 0.1cm
\begin{subfigure}[b]{0.3\linewidth}
\centering
\includegraphics[width=1.0\linewidth]{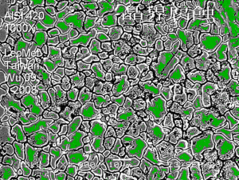}
\end{subfigure}
\hskip 0.1cm
\begin{subfigure}[b]{0.3\linewidth}
\centering
\includegraphics[width=1.0\linewidth]{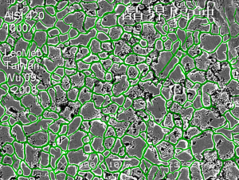}
\end{subfigure}

\vskip 0.1cm
\begin{subfigure}[b]{0.3\linewidth}
\centering
\includegraphics[width=1.0\linewidth]{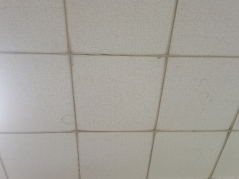}
\end{subfigure}
\hskip 0.1cm
\begin{subfigure}[b]{0.3\linewidth}
\centering
\includegraphics[width=1.0\linewidth]{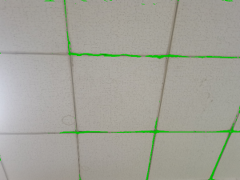}
\end{subfigure}
\hskip 0.1cm
\begin{subfigure}[b]{0.3\linewidth}
\centering
\includegraphics[width=1.0\linewidth]{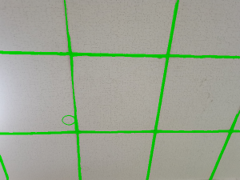}
\end{subfigure}

\vskip 0.1cm
\begin{subfigure}[b]{0.3\linewidth}
\centering
\includegraphics[width=1.0\linewidth]{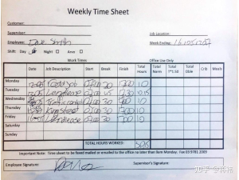}
\end{subfigure}
\hskip 0.1cm
\begin{subfigure}[b]{0.3\linewidth}
\centering
\includegraphics[width=1.0\linewidth]{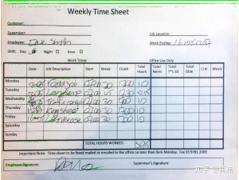}
\end{subfigure}
\hskip 0.1cm
\begin{subfigure}[b]{0.3\linewidth}
\centering
\includegraphics[width=1.0\linewidth]{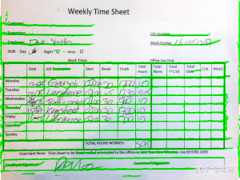}
\end{subfigure}

\vskip 0.1cm
\begin{subfigure}[b]{0.3\linewidth}
\centering
\includegraphics[width=1.0\linewidth]{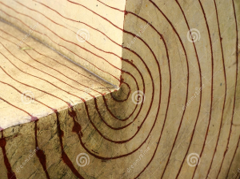}
\end{subfigure}
\hskip 0.1cm
\begin{subfigure}[b]{0.3\linewidth}
\centering
\includegraphics[width=1.0\linewidth]{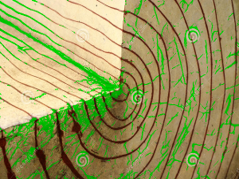}
\end{subfigure}
\hskip 0.1cm
\begin{subfigure}[b]{0.3\linewidth}
\centering
\includegraphics[width=1.0\linewidth]{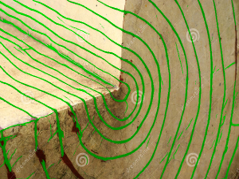}
\end{subfigure}

\vskip 0.1cm
\begin{subfigure}[b]{0.3\linewidth}
\centering
\includegraphics[width=1.0\linewidth]{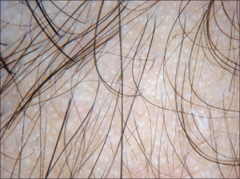}
\end{subfigure}
\hskip 0.1cm
\begin{subfigure}[b]{0.3\linewidth}
\centering
\includegraphics[width=1.0\linewidth]{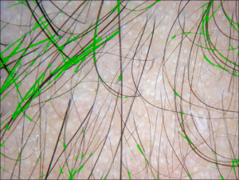}
\end{subfigure}
\hskip 0.1cm
\begin{subfigure}[b]{0.3\linewidth}
\centering
\includegraphics[width=1.0\linewidth]{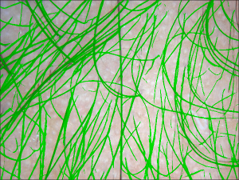}
\end{subfigure}

\caption{Some segmentation results on different types of curvilinear images using the model trained on DRIVE dataset. From left to right: input image, segmentation on the original image using the baseline model, segmentation result by LIOT. Best viewed by zooming in the electronic version.}
\label{fig:generalization_type}
\end{figure}

To further demonstrate the generalizability of the proposed LIOT, we also apply the model trained on the DRIVE dataset to images with different types of curvilinear structures: material grains, ceiling grids, table lines, tree's growth rings, and hair. As illustrated in Fig.~\ref{fig:generalization_type}, though the adopted model is trained on the retinal images having very different appearances from these test images, LIOT can still successfully segment these curvilinear structures of different types. Whereas, the baseline model trained on the original images fails to extract many curvilinear structures and has many false positives. This illustrative experiment further implies that LIOT is able to capture the inherent characteristic of curvilinear objects, and can thus improve the generalizability of existing deep learning-based models for curvilinear object segmentation. 


\section{Conclusion}\label{sec:conclusion}
In this paper, we aim to improve the generalization ability of current deep learning-based curvilinear object segmentation methods. For that, we propose the LIOT that
converts a gray-scale image to a novel representation that is
invariant to increasing contrast changes. LIOT is built on the intensity order between pairs of pixels, and thus does not depend on
the absolute intensity values. Such intensity-order based
representation captures the inherent property of curvilinear objects
(\eg, curvilinear structure darker than the context). Despite its simplicity, LIOT is rather robust to non-targeted universal adversarial perturbations. Besides, extensive cross-dataset
experiments on three widely adopted retinal blood vessel segmentation
datasets and CrackTree dataset demonstrate that LIOT can improve the classical segmentation pipeline that
directly operates on the original image. 
Therefore, LIOT forms a simple yet effective way to improve the generalization performance of
different models. To the best of our knowledge, LIOT is a first simple attempt to improve the generalizability of curvilinear object segmentation from the aspect of image transformation. More sophisticated image transformation is worthy of further investigation for robust curvilinear object segmentation.

\bibliographystyle{IEEEtran}

\bibliography{reference}

\begin{thebibliography}{10}
\providecommand{\url}[1]{#1}
\csname url@samestyle\endcsname
\providecommand{\newblock}{\relax}
\providecommand{\bibinfo}[2]{#2}
\providecommand{\BIBentrySTDinterwordspacing}{\spaceskip=0pt\relax}
\providecommand{\BIBentryALTinterwordstretchfactor}{4}
\providecommand{\BIBentryALTinterwordspacing}{\spaceskip=\fontdimen2\font plus
\BIBentryALTinterwordstretchfactor\fontdimen3\font minus
  \fontdimen4\font\relax}
\providecommand{\BIBforeignlanguage}[2]{{%
\expandafter\ifx\csname l@#1\endcsname\relax
\typeout{** WARNING: IEEEtran.bst: No hyphenation pattern has been}%
\typeout{** loaded for the language `#1'. Using the pattern for}%
\typeout{** the default language instead.}%
\else
\language=\csname l@#1\endcsname
\fi
#2}}
\providecommand{\BIBdecl}{\relax}
\BIBdecl

\bibitem{abramoff2010retinal}
M.~D. Abr{\`a}moff, M.~K. Garvin, and M.~Sonka, ``Retinal imaging and image
  analysis,'' \emph{IEEE Reviews in Biomedical Engineering}, vol.~3, pp.
  169--208, 2010.

\bibitem{fraz2012blood}
M.~M. Fraz, P.~Remagnino, A.~Hoppe, B.~Uyyanonvara, A.~R. Rudnicka, C.~G. Owen,
  and S.~A. Barman, ``Blood vessel segmentation methodologies in retinal
  images--a survey,'' \emph{Computer Methods and Programs in Biomedicine}, vol.
  108, no.~1, pp. 407--433, 2012.

\bibitem{liu2011fingerprint}
E.~Liu, H.~Zhao, F.~Guo, J.~Liang, and J.~Tian, ``Fingerprint segmentation
  based on an adaboost classifier,'' \emph{Frontiers of Computer Science},
  vol.~5, no.~2, pp. 148--157, 2011.

\bibitem{laibacher2019m2u}
T.~Laibacher, T.~Weyde, and S.~Jalali, ``{M2U-Net}: Effective and efficient
  retinal vessel segmentation for real-world applications,'' in \emph{Proc. of
  IEEE Intl. Conf. on Computer Vision and Pattern Recognition Workshops}, 2019,
  pp. 0--0.

\bibitem{su2019non}
K.~Su, G.~Yang, L.~Yang, P.~Su, and Y.~Yin, ``Non-negative locality-constrained
  vocabulary tree for finger vein image retrieval,'' \emph{Frontiers of
  Computer Science}, vol.~13, no.~2, pp. 318--332, 2019.

\bibitem{zou2012cracktree}
Q.~Zou, Y.~Cao, Q.~Li, Q.~Mao, and S.~Wang, ``Cracktree: Automatic crack
  detection from pavement images,'' \emph{Pattern Recognition Letters},
  vol.~33, no.~3, pp. 227--238, 2012.

\bibitem{bibiloni2016survey}
P.~Bibiloni, M.~Gonz{\'a}lez-Hidalgo, and S.~Massanet, ``A survey on
  curvilinear object segmentation in multiple applications,'' \emph{Pattern
  Recognition}, vol.~60, pp. 949--970, 2016.

\bibitem{koller1995multiscale}
T.~M. Koller, G.~Gerig, G.~Szekely, and D.~Dettwiler, ``Multiscale detection of
  curvilinear structures in {2-D} and {3-D} image data,'' in \emph{Proc. of
  IEEE Intl. Conf. on Computer Vision}, 1995, pp. 864--869.

\bibitem{carlotto2006enhancement}
M.~J. Carlotto, ``Enhancement of low-contrast curvilinear features in
  imagery,'' \emph{IEEE Trans. on Image Processing}, vol.~16, no.~1, pp.
  221--228, 2006.

\bibitem{soares2006Gabor_waveletretinal}
J.~V. Soares, J.~J. Leandro, R.~M. Cesar, H.~F. Jelinek, and M.~J. Cree,
  ``Retinal vessel segmentation using the {2-D Gabor} wavelet and supervised
  classification,'' \emph{IEEE Trans. on Medical Imaging}, vol.~25, no.~9, pp.
  1214--1222, 2006.

\bibitem{ricci2007linedetector}
E.~Ricci and R.~Perfetti, ``Retinal blood vessel segmentation using line
  operators and support vector classification,'' \emph{IEEE Trans. on Medical
  Imaging}, vol.~26, no.~10, pp. 1357--1365, 2007.

\bibitem{al2009active}
B.~Al-Diri, A.~Hunter, and D.~Steel, ``An active contour model for segmenting
  and measuring retinal vessels,'' \emph{IEEE Trans. on Medical Imaging},
  vol.~28, no.~9, pp. 1488--1497, 2009.

\bibitem{marin2010new}
D.~Mar{\'\i}n, A.~Aquino, M.~E. Geg{\'u}ndez-Arias, and J.~M. Bravo, ``A new
  supervised method for blood vessel segmentation in retinal images by using
  gray-level and moment invariants-based features,'' \emph{IEEE Trans. on
  Medical Imaging}, vol.~30, no.~1, pp. 146--158, 2010.

\bibitem{obara2012contrast}
B.~Obara, M.~Fricker, D.~Gavaghan, and V.~Grau, ``Contrast-independent
  curvilinear structure detection in biomedical images,'' \emph{IEEE Trans. on
  Image Processing}, vol.~21, no.~5, pp. 2572--2581, 2012.

\bibitem{xiao2012multiscale}
C.~Xiao, M.~Staring, Y.~Wang, D.~P. Shamonin, and B.~C. Stoel, ``Multiscale
  {bi-Gaussian} filter for adjacent curvilinear structures detection with
  application to vasculature images,'' \emph{IEEE Trans. on Image Processing},
  vol.~22, no.~1, pp. 174--188, 2012.

\bibitem{krylov2014stochastic}
V.~A. Krylov and J.~D. Nelson, ``Stochastic extraction of elongated curvilinear
  structures with applications,'' \emph{IEEE Trans. on Image Processing},
  vol.~23, no.~12, pp. 5360--5373, 2014.

\bibitem{annunziata2015boosting}
R.~Annunziata, A.~Kheirkhah, P.~Hamrah, and E.~Trucco, ``Boosting hand-crafted
  features for curvilinear structure segmentation by learning context
  filters,'' in \emph{Proc. of Intl. Conf. on Medical Image Computing and
  Computer Assisted Intervention}, 2015, pp. 596--603.

\bibitem{vicas2015detecting}
C.~Vicas and S.~Nedevschi, ``Detecting curvilinear features using structure
  tensors,'' \emph{IEEE Trans. on Image Processing}, vol.~24, no.~11, pp.
  3874--3887, 2015.

\bibitem{turetken2016reconstructing}
E.~T{\"u}retken, F.~Benmansour, B.~Andres, P.~G{\l}owacki, H.~Pfister, and
  P.~Fua, ``Reconstructing curvilinear networks using path classifiers and
  integer programming,'' \emph{IEEE Trans. Pattern Anal. Mach. Intell.},
  vol.~38, no.~12, pp. 2515--2530, 2016.

\bibitem{merveille2017curvilinear}
O.~Merveille, H.~Talbot, L.~Najman, and N.~Passat, ``Curvilinear structure
  analysis by ranking the orientation responses of path operators,'' \emph{IEEE
  Trans. Pattern Anal. Mach. Intell.}, vol.~40, no.~2, pp. 304--317, 2017.

\bibitem{strisciuglio2019robust}
N.~Strisciuglio, G.~Azzopardi, and N.~Petkov, ``Robust inhibition-augmented
  operator for delineation of curvilinear structures,'' \emph{IEEE Trans. on
  Image Processing}, vol.~28, no.~12, pp. 5852--5866, 2019.

\bibitem{merveille2019n}
O.~Merveille, B.~Naegel, H.~Talbot, and N.~Passat, ``{$ n $ D} variational
  restoration of curvilinear structures with prior-based directional
  regularization,'' \emph{IEEE Trans. on Image Processing}, vol.~28, no.~8, pp.
  3848--3859, 2019.

\bibitem{li2019iternet}
L.~Li, M.~Verma, Y.~Nakashima, H.~Nagahara, and R.~Kawasaki, ``{IterNet}:
  Retinal image segmentation utilizing structural redundancy in vessel
  networks,'' in \emph{Proc. of IEEE Winter Conf. on Applications of Computer
  Vision}, 2020, pp. 3656--3665.

\bibitem{li2015cross}
Q.~Li, B.~Feng, L.~Xie, P.~Liang, H.~Zhang, and T.~Wang, ``A cross-modality
  learning approach for vessel segmentation in retinal images,'' \emph{IEEE
  Trans. on Medical Imaging}, vol.~35, no.~1, pp. 109--118, 2015.

\bibitem{maninis2016DRIU}
K.-K. Maninis, J.~Pont-Tuset, P.~Arbel{\'a}ez, and L.~Van~Gool, ``Deep retinal
  image understanding,'' in \emph{Proc. of Intl. Conf. on Medical Image
  Computing and Computer Assisted Intervention}, 2016, pp. 140--148.

\bibitem{mosinska2018beyond}
A.~Mosinska, P.~Marquez-Neila, M.~Kozi{\'n}ski, and P.~Fua, ``Beyond the
  pixel-wise loss for topology-aware delineation,'' in \emph{Proc. of IEEE
  Intl. Conf. on Computer Vision and Pattern Recognition}, 2018, pp.
  3136--3145.

\bibitem{yan2018joint}
Z.~Yan, X.~Yang, and K.-T. Cheng, ``Joint segment-level and pixel-wise losses
  for deep learning based retinal vessel segmentation,'' \emph{IEEE
  Transactions on Biomedical Engineering}, vol.~65, no.~9, pp. 1912--1923,
  2018.

\bibitem{luan2018gabor}
S.~Luan, C.~Chen, B.~Zhang, J.~Han, and J.~Liu, ``Gabor convolutional
  networks,'' \emph{IEEE Trans. on Image Processing}, vol.~27, no.~9, pp.
  4357--4366, 2018.

\bibitem{mosinska2019joint}
A.~Mosinska, M.~Kozi{\'n}ski, and P.~Fua, ``Joint segmentation and path
  classification of curvilinear structures,'' \emph{IEEE Trans. Pattern Anal.
  Mach. Intell.}, vol.~42, no.~6, pp. 1515--1521, 2019.

\bibitem{wang2019context}
F.~Wang, Y.~Gu, W.~Liu, Y.~Yu, S.~He, and J.~Pan, ``Context-aware
  spatio-recurrent curvilinear structure segmentation,'' in \emph{Proc. of IEEE
  Intl. Conf. on Computer Vision and Pattern Recognition}, 2019, pp.
  12\,648--12\,657.

\bibitem{mou2019cs}
L.~Mou, Y.~Zhao, L.~Chen, J.~Cheng, Z.~Gu, H.~Hao, H.~Qi, Y.~Zheng, A.~Frangi,
  and J.~Liu, ``{CS-Net}: channel and spatial attention network for curvilinear
  structure segmentation,'' in \emph{Proc. of Intl. Conf. on Medical Image
  Computing and Computer Assisted Intervention}, 2019, pp. 721--730.

\bibitem{guo2019bts}
S.~Guo, K.~Wang, H.~Kang, Y.~Zhang, Y.~Gao, and T.~Li, ``{BTS-DSN}: Deeply
  supervised neural network with short connections for retinal vessel
  segmentation,'' \emph{International Journal of Medical Informatics}, vol.
  126, pp. 105--113, 2019.

\bibitem{cherukuri2019deep}
V.~Cherukuri, V.~K. BG, R.~Bala, and V.~Monga, ``Deep retinal image
  segmentation with regularization under geometric priors,'' \emph{IEEE Trans.
  on Image Processing}, vol.~29, pp. 2552--2567, 2019.

\bibitem{dey2020subpixel}
S.~Dey, ``A subpixel residual {U-Net} and feature fusion preprocessing for
  retinal vessel segmentation,'' in \emph{Proc. of European Conference on
  Computer Vision}, 2020, pp. 239--250.

\bibitem{mou2020cs2}
L.~Mou, Y.~Zhao, H.~Fu, Y.~Liu, J.~Cheng, Y.~Zheng, P.~Su, J.~Yang, L.~Chen,
  A.~F. Frangi \emph{et~al.}, ``{CS2-Net}: Deep learning segmentation of
  curvilinear structures in medical imaging,'' \emph{Medical Image Analysis},
  vol.~67, p. 101874, 2020.

\bibitem{ding2020novel}
L.~Ding, M.~H. Bawany, A.~E. Kuriyan, R.~S. Ramchandran, C.~C. Wykoff, and
  G.~Sharma, ``A novel deep learning pipeline for retinal vessel detection in
  fluorescein angiography,'' \emph{IEEE Trans. on Image Processing}, vol.~29,
  pp. 6561--6573, 2020.

\bibitem{zhang20203d}
J.~Zhang, Y.~Qiao, M.~S. Sarabi, M.~M. Khansari, J.~K. Gahm, A.~H. Kashani, and
  Y.~Shi, ``{3D} shape modeling and analysis of retinal microvasculature in
  {OCT}-angiography images,'' \emph{IEEE Trans. on Medical Imaging}, vol.~39,
  no.~5, p. 1335, 2020.

\bibitem{nazir2020off}
A.~Nazir, M.~N. Cheema, B.~Sheng, H.~Li, P.~Li, P.~Yang, Y.~Jung, J.~Qin,
  J.~Kim, and D.~D. Feng, ``{OFF-eNET}: An optimally fused fully end-to-end
  network for automatic dense volumetric {3D} intracranial blood vessels
  segmentation,'' \emph{IEEE Trans. on Image Processing}, vol.~29, pp.
  7192--7202, 2020.

\bibitem{wu2021scs}
H.~Wu, W.~Wang, J.~Zhong, B.~Lei, Z.~Wen, and J.~Qin, ``{SCS-Net}: A scale and
  context sensitive network for retinal vessel segmentation,'' \emph{Medical
  Image Analysis}, vol.~70, p. 102025, 2021.

\bibitem{shen2021modeling}
Z.~Shen, H.~Fu, J.~Shen, and L.~Shao, ``Modeling and enhancing low-quality
  retinal fundus images,'' \emph{IEEE Trans. on Medical Imaging}, vol.~40,
  no.~3, pp. 996--1006, 2021.

\bibitem{lecun2015deep}
Y.~LeCun, Y.~Bengio, and G.~Hinton, ``Deep learning,'' \emph{nature}, vol. 521,
  no. 7553, pp. 436--444, 2015.

\bibitem{he2020hybrid}
N.~He, L.~Fang, and A.~Plaza, ``Hybrid first and second order attention {U}net
  for building segmentation in remote sensing images,'' \emph{Science China
  Information Sciences}, vol.~63, no.~4, pp. 1--12, 2020.

\bibitem{yue2021self}
J.~Yue, L.~Fang, H.~Rahmani, and P.~Ghamisi, ``Self-supervised learning with
  adaptive distillation for hyperspectral image classification,'' \emph{IEEE
  Transactions on Geoscience and Remote Sensing}, vol.~60, 2021.

\bibitem{xie2021super}
J.~Xie, L.~Fang, B.~Zhang, J.~Chanussot, and S.~Li, ``Super resolution guided
  deep network for land cover classification from remote sensing images,''
  \emph{IEEE Transactions on Geoscience and Remote Sensing}, 2021.

\bibitem{DRIVE2004ridge}
J.~Staal, M.~D. Abr{\`a}moff, M.~Niemeijer, M.~A. Viergever, and
  B.~Van~Ginneken, ``Ridge-based vessel segmentation in color images of the
  retina,'' \emph{IEEE Trans. on Medical Imaging}, vol.~23, no.~4, pp.
  501--509, 2004.

\bibitem{STARE2000locating}
A.~Hoover, V.~Kouznetsova, and M.~Goldbaum, ``Locating blood vessels in retinal
  images by piecewise threshold probing of a matched filter response,''
  \emph{IEEE Trans. on Medical Imaging}, vol.~19, no.~3, pp. 203--210, 2000.

\bibitem{fraz2012ensemble}
M.~M. Fraz, P.~Remagnino, A.~Hoppe, B.~Uyyanonvara, A.~R. Rudnicka, C.~G. Owen,
  and S.~A. Barman, ``An ensemble classification-based approach applied to
  retinal blood vessel segmentation,'' \emph{IEEE Transactions on Biomedical
  Engineering}, vol.~59, no.~9, pp. 2538--2548, 2012.

\bibitem{wang2020higher}
J.~Wang and A.~C. Chung, ``Higher-order flux with spherical harmonics transform
  for vascular analysis,'' in \emph{Proc. of Intl. Conf. on Medical Image
  Computing and Computer Assisted Intervention}.\hskip 1em plus 0.5em minus
  0.4em\relax Springer, 2020, pp. 55--65.

\bibitem{ma2021rose}
Y.~Ma, H.~Hao, J.~Xie, H.~Fu, J.~Zhang, J.~Yang, Z.~Wang, J.~Liu, Y.~Zheng, and
  Y.~Zhao, ``{ROSE}: A retinal oct-angiography vessel segmentation dataset and
  new model,'' \emph{IEEE Trans. on Medical Imaging}, vol.~40, no.~3, pp.
  928--939, 2021.

\bibitem{wang2019tubular}
C.~Wang, Y.~Hayashi, M.~Oda, H.~Itoh, T.~Kitasaka, A.~F. Frangi, and K.~Mori,
  ``Tubular structure segmentation using spatial fully connected network with
  radial distance loss for {3D} medical images,'' in \emph{Proc. of Intl. Conf.
  on Medical Image Computing and Computer Assisted Intervention}, 2019, pp.
  348--356.

\bibitem{wang2020deep}
Y.~Wang, X.~Wei, F.~Liu, J.~Chen, Y.~Zhou, W.~Shen, E.~K. Fishman, and A.~L.
  Yuille, ``Deep distance transform for tubular structure segmentation in {CT}
  scans,'' in \emph{Proc. of IEEE Intl. Conf. on Computer Vision and Pattern
  Recognition}, 2020, pp. 3833--3842.

\bibitem{hu2019topology}
X.~Hu, L.~Fuxin, D.~Samaras, and C.~Chen, ``Topology-preserving deep image
  segmentation,'' in \emph{Proc. of Advances in Neural Information Processing
  Systems}, 2019.

\bibitem{hu2021topology}
X.~Hu, Y.~Wang, L.~Fuxin, D.~Samaras, and C.~Chen, ``Topology-aware
  segmentation using discrete morse theory,'' in \emph{Proc. of International
  Conference on Learning Representations}, 2021.

\bibitem{shit2021cldice}
S.~Shit, J.~C. Paetzold, A.~Sekuboyina, I.~Ezhov, A.~Unger, A.~Zhylka, J.~P.
  Pluim, U.~Bauer, and B.~H. Menze, ``{clDice}-a novel topology-preserving loss
  function for tubular structure segmentation,'' in \emph{Proc. of IEEE Intl.
  Conf. on Computer Vision and Pattern Recognition}, 2021, pp.
  16\,560--16\,569.

\bibitem{wang2021single}
H.~Wang, Y.~Song, C.~Zhang, J.~Yu, S.~Liu, H.~Pengy, and W.~Cai, ``Single
  neuron segmentation using graph-based global reasoning with auxiliary
  skeleton loss from {3D} optical microscope images,'' in \emph{Proc. of Intl.
  Symp. on Biomedical Imaging}, 2021, pp. 934--938.

\bibitem{cheng2021joint}
M.~Cheng, K.~Zhao, X.~Guo, Y.~Xu, and J.~Guo, ``Joint topology-preserving and
  feature-refinement network for curvilinear structure segmentation,'' in
  \emph{Proc. of IEEE Intl. Conf. on Computer Vision}, 2021, pp. 7147--7156.

\bibitem{zabih1996non}
R.~Zabih and J.~Woodfill, ``A non-parametric approach to visual
  correspondence,'' in \emph{IEEE Trans. Pattern Anal. Mach. Intell.}, 1996.

\bibitem{chang2010algorithm}
N.~Y.-C. Chang, T.-H. Tsai, B.-H. Hsu, Y.-C. Chen, and T.-S. Chang, ``Algorithm
  and architecture of disparity estimation with mini-census adaptive support
  weight,'' \emph{IEEE Transactions on Circuits and Systems for Video
  Technology}, vol.~20, no.~6, pp. 792--805, 2010.

\bibitem{lee2016improved}
J.~Lee, D.~Jun, C.~Eem, and H.~Hong, ``Improved census transform for noise
  robust stereo matching,'' \emph{Optical Engineering}, vol.~55, no.~6, p.
  063107, 2016.

\bibitem{froba2004face}
B.~Froba and A.~Ernst, ``Face detection with the modified census transform,''
  in \emph{IEEE International Conference on Automatic Face and Gesture
  Recognition}, 2004, pp. 91--96.

\bibitem{ambrosch2010miniature}
K.~Ambrosch, C.~Zinner, and H.~Leopold, ``A miniature embedded stereo vision
  system for automotive applications,'' in \emph{IEEE Convention of Electrical
  and Electronics Engineers in Israel}, 2010, pp. 000\,786--000\,789.

\bibitem{fife2012improved}
W.~S. Fife and J.~K. Archibald, ``Improved census transforms for
  resource-optimized stereo vision,'' \emph{IEEE Transactions on Circuits and
  Systems for Video Technology}, vol.~23, no.~1, pp. 60--73, 2012.

\bibitem{spangenberg2013weighted}
R.~Spangenberg, T.~Langner, and R.~Rojas, ``Weighted semi-global matching and
  center-symmetric census transform for robust driver assistance,'' in
  \emph{International Conference on Computer Analysis of Images and Patterns},
  2013, pp. 34--41.

\bibitem{ahlberg2019unbounded}
C.~Ahlberg, M.~L. Ortiz, F.~Ekstrand, and M.~Ekstrom, ``Unbounded sparse census
  transform using genetic algorithm,'' in \emph{Proc. of IEEE Winter Conf. on
  Applications of Computer Vision}, 2019, pp. 1616--1625.

\bibitem{yu2020stereo}
S.~Yu, Y.~He, Z.~Chen, C.~Ru, and M.~Pang, ``Stereo matching method based on
  combination characteristic cost computing and unstable tree reconstruction
  optimization and its application in medical images,'' \emph{Journal of
  Medical Imaging and Health Informatics}, vol.~10, no.~3, pp. 646--653, 2020.

\bibitem{lai2019efficient}
X.~Lai, X.~Xu, J.~Zhang, Y.~Fang, and Z.~Huang, ``An efficient implementation
  of a census-based stereo matching and its applications in medical imaging,''
  \emph{Journal of Medical Imaging and Health Informatics}, vol.~9, no.~6, pp.
  1152--1159, 2019.

\bibitem{cornejo2019audio}
J.~Y.~R. Cornejo and H.~Pedrini, ``Audio-visual emotion recognition using a
  hybrid deep convolutional neural network based on census transform,'' in
  \emph{IEEE International Conference on Systems, Man and Cybernetics}, 2019,
  pp. 3396--3402.

\bibitem{wang2011local}
Z.~Wang, B.~Fan, and F.~Wu, ``Local intensity order pattern for feature
  description,'' in \emph{Proc. of IEEE Intl. Conf. on Computer Vision}.\hskip
  1em plus 0.5em minus 0.4em\relax IEEE, 2011, pp. 603--610.

\bibitem{wang2015exploring}
Z.~Wang, B.~Fan, G.~Wang, and F.~Wu, ``Exploring local and overall ordinal
  information for robust feature description,'' \emph{IEEE Trans. Pattern Anal.
  Mach. Intell.}, vol.~38, no.~11, pp. 2198--2211, 2015.

\bibitem{fan2011rotationally}
B.~Fan, F.~Wu, and Z.~Hu, ``Rotationally invariant descriptors using intensity
  order pooling,'' \emph{IEEE Trans. Pattern Anal. Mach. Intell.}, vol.~34,
  no.~10, pp. 2031--2045, 2011.

\bibitem{shen2014hep}
L.~Shen, J.~Lin, S.~Wu, and S.~Yu, ``{HEp-2} image classification using
  intensity order pooling based features and bag of words,'' \emph{Pattern
  Recognition}, vol.~47, no.~7, pp. 2419--2427, 2014.

\bibitem{lei2014local}
Z.~Lei, D.~Yi, and S.~Z. Li, ``Local gradient order pattern for face
  representation and recognition,'' in \emph{Proc. of Intl. Conf. on Pattern
  Recognition}.\hskip 1em plus 0.5em minus 0.4em\relax IEEE, 2014, pp.
  387--392.

\bibitem{wang2017ordinal}
G.~Wang, B.~Fan, Z.~Zhou, and C.~Pan, ``Ordinal pyramid coding for rotation
  invariant feature extraction,'' \emph{Neurocomputing}, vol. 242, pp.
  150--160, 2017.

\bibitem{evaluation2011function}
M.~E. Geg{\'u}ndez-Arias, A.~Aquino, J.~M. Bravo, and D.~Mar{\'\i}n, ``A
  function for quality evaluation of retinal vessel segmentations,'' \emph{IEEE
  Trans. on Medical Imaging}, vol.~31, no.~2, pp. 231--239, 2011.

\bibitem{kingma2014adam}
D.~P. Kingma and J.~Ba, ``Adam: A method for stochastic optimization,'' in
  \emph{Proc. of International Conference on Learning Representations}, 2015.

\bibitem{poursaeed2018generative}
O.~Poursaeed, I.~Katsman, B.~Gao, and S.~Belongie, ``Generative adversarial
  perturbations,'' in \emph{Proc. of IEEE Intl. Conf. on Computer Vision and
  Pattern Recognition}, 2018, pp. 4422--4431.

\bibitem{oliveira2018retinal}
A.~Oliveira, S.~Pereira, and C.~A. Silva, ``Retinal vessel segmentation based
  on fully convolutional neural networks,'' \emph{Expert Systems with
  Applications}, vol. 112, pp. 229--242, 2018.

\bibitem{torp2018temporal}
T.~L. Torp, R.~Kawasaki, T.~Y. Wong, T.~Peto, and J.~Grauslund, ``Temporal
  changes in retinal vascular parameters associated with successful panretinal
  photocoagulation in proliferative diabetic retinopathy: a prospective
  clinical interventional study,'' \emph{Acta Ophthalmologica}, vol.~96, no.~4,
  pp. 405--410, 2018.

\end{thebibliography}

\end{document}